\documentclass[aps,prb,showpacs,twocolumn,floatfix]{revtex4}

\usepackage{amsmath}
\usepackage{graphicx}
\usepackage{dcolumn}
\usepackage{bm}

\newcommand{\kb}{k_{\rm B}}

\renewcommand{\Re}{\operatorname{Re}}

\begin{document}
\title{Stochastic Transition States: Reaction Geometry amidst Noise}
\author{Thomas Bartsch}
\affiliation{Center for Nonlinear Science and School of Physics,
  Georgia Institute of Technology, Atlanta, GA 30332-0430, USA}
\author{Rigoberto Hernandez}
\affiliation{Center for Nonlinear Science and
             School of Chemistry and Biochemistry,
             Georgia Institute of Technology, Atlanta, GA 30332-0300, USA}
\author{T. Uzer}
\affiliation{Center for Nonlinear Science and School of Physics,
  Georgia Institute of Technology, Atlanta, GA 30332-0430, USA}
\date{\today}

\begin{abstract}
Classical transition state theory (TST) is the cornerstone of reaction rate
theory. It postulates a partition of phase space into reactant and product
regions, which are separated by a dividing surface that reactive
trajectories must cross. 
In order not to overestimate the reaction rate,
the dynamics must be free of recrossings of the dividing surface.
This no-recrossing rule is 
difficult (and sometimes impossible) to enforce, 
however, when a chemical reaction takes place in a
fluctuating environment such as a liquid.
High-accuracy approximations to the rate are well known when 
the solvent forces are treated using stochastic representations, though
again, exact no-recrossing surfaces have not been
available.
To generalize the exact limit of TST to reactive
systems driven by noise, we introduce a time-dependent dividing surface
that is stochastically moving in phase space such that it is crossed 
once and only once by each transition path.
\end{abstract}
\pacs{82.20.Db, 34.10.+x, , 05.40.Ca, 05.45.-a}
\maketitle

\section{Introduction}
\label{sec:Intro}

Transition state theory (TST) 
\cite{eyring35,Wigner38,Wigner39,Miller,Keck67,Truhlar96,Jaffe05} 
plays a central role in chemistry because it provides both a tantalizingly
simple
approach to calculating chemical reaction rates and 
an intuitive interpretation of reaction mechanisms 
through the identification
of the bottleneck between reactants and products.
Although the rate could in principle be calculated
exactly using known methods, in practice, it is often out of reach
because such computations are laborious, 
even if quantum effects are ignored.\cite{goldberger,baer85,mill93}
Therefore, the TST approximation to the rate has played a central
role in kinetics, but its accuracy
hinges on the optimal identification of 
the TS dividing surface.\cite{pech73b}
Indeed TST is exact when the latter 
is crossed once and only once by each reactive trajectory.\cite{Keck67,truh84}
While a constructive prescription for a no-recrossing 
TS dividing surface has long been sought, it was only in the 70's
that it was found in the special case of
low-dimensional systems.\cite{pech73b,pech79b,pollak85} 
A general method that
yields a no-recrossing TS dividing surface for reactions
in arbitrarily many (but finite)
degrees of freedom has been described only recently.
\cite{Komatsuzaki99,bristol1,bristol2,Jaffe05}
Nonetheless, even approximate TS dividing surfaces located 
near the phase-space bottlenecks between reactants and products 
provide a portrait of the activated complex 
that has been used routinely in the chemistry community. 

Reactions in solution are even more complex: 
the presence of the solvent introduces
a complicated many-body problem.
Nevertheless, TST has been remarkably successful at
providing accurate estimates of thermal reaction rates when a reasonable
TS dividing surface can be identified.
\cite{hynes85b,pollak87,berne88,pollak96}
Unfortunately, such estimates are seldom exact because the dividing
surfaces available in the literature are not strictly free of recrossings.
One approach lies in choosing the dividing surface 
so as to minimize the TST reaction rate.
\cite{truh84,pollak90a,frish92,tucker92,tucker95}
The recrossing problem is thereby alleviated, but usually not eliminated.
Alternatively, one can eliminate consideration of
the nonreactive trajectories entirely
by sampling the transition path ensemble (TPE)
directly.\cite{chan98a,chandler99b,chandler00,chan02a,chan02b}
All such paths intersect the TS dividing surface and hence,
as a collection, they provide a portrait of the activated complex.
Using Monte Carlo approaches,\cite{chan02a,chan02b}
TPE can also be readily applied to 
complex molecular reactions in all-atom solvents
but it does not provide the simple geometric picture of a
dividing surface free of recrossings.

The geometry of the finite dimensional no-recrossing TS dividing surface 
is described in this paper for chemical reactions that can
be represented using stochastic models of the solvent interactions.
The moving TS dividing surface is most readily constructed in cases of uniform
solvents such as those represented by the Langevin equation
as was summarized in a recent Letter.\cite{dawn05}
The present article expands that discussion, and extends the results
to the colored noise case of the generalized Langevin equation.
It should be noted that a finite-dimensional-like TS dividing surface 
can be specified if an explicit infinite-dimensional model
---{\it viz.}~an infinite collection of harmonic oscillators---
is used to represent the heat bath 
in the Langevin equation.\cite{zwan73,caldeira81,pollak86b} 
In spite of its simplicity, this model leads to an excellent
approximation to the rate constant\cite{pollak86b,pgh89} that, remarkably,
has been rederived without recourse to an explicit model of the heat 
bath.\cite{Graham90}
Within the finite-dimensional phase space of the solute, the
reaction geometry has also been illustrated 
using the stochastic
separatrix (i.e., the collection of all phase space points for which the
reaction probability is 50\%.) 
That work portends the present development.\cite{Martens02} 

In this paper, we introduce a generalized view of the
transition state that allows for the temporal variation of the solvent.
While this has not yet been accomplished in an all-atom representation, we
show here that if the solvent is modeled through the Langevin equation, a
time-dependent TS dividing surface can be constructed that is
strictly free of recrossings. This is achieved by identifying a privileged
stochastic trajectory that remains in the vicinity of the barrier for all
time. This Transition State Trajectory serves as a moving origin to which
geometric structures in a noiseless phase space, such as invariant manifolds
and a no-recrossing surface\cite{bristol1,bristol2,Jaffe05} are attached.
The time-dependent dividing surface thus obtained
adds to the evolving understanding of the geometric structures
separating reactants from products in the presence of noise.
The picture presented here provides a detailed time-resolved description of
the reaction dynamics and, in particular, gives a precise geometric
interpretation of the collective reaction coordinate introduced 
in Ref.~\onlinecite{Graham90}.

The outline of the present paper is as follows: Section~\ref{sec:prelim}
recapitulates the macroscopic model of the solvent dynamics represented by the
Langevin equation. In section~\ref{sec:fund}, we
introduce the fundamental construction that allows us to separate the full
dynamics into the noisy TS trajectory and the noiseless
relative motion. 
The TS trajectory for a particle driven by a white noise bath
is described explicitly in section~\ref{sec:WN}.
The construction is simple and direct, and requires no more effort than that
for a typical trajectory.
Several statistical properties of the TS trajectory are also 
discussed in section~\ref{sec:WN}.
In addition, the TS trajectory is used to
to compute reaction probabilities and the stochastic separatrix. 
The generalization of the TS trajectory to 
the chemically important case of colored noise
is discussed in section~\ref{sec:Coloured}.

\section{Preliminaries}
\label{sec:prelim}

The influence of the solvent on a chemical motion or reaction coordinate,
$\vec q$, has been routinely represented using the 
Langevin equation (LE) of motion,\cite{rmp90,zwan01}
\begin{equation}
  \label{LE}
  \ddot{\vec q}_\alpha(t) = -\nabla_{\vec q} U(\vec q_\alpha(t)) -
        \bm{\Gamma}\dot{\vec q}_\alpha(t) +
        \vec\xi_\alpha(t)\;,
\end{equation}
or its generalization (GLE) to allow for colored 
noise.\cite{kram40,zwan61a,zwan61b,prig61,mori65,fox78,rmp90,zwan01}
In Eq.~\ref{LE}, the vector $\vec q$ 
denotes a set of mass-weighted coordinates in
a configuration space of arbitrary dimension $N$, 
$U(\vec q)$ the potential of mean force governing the reaction, 
$\bm\Gamma$ a symmetric positive-definite friction matrix and 
$\vec\xi_\alpha(t)$ a stochastic force. 
The subscript $\alpha$ here and throughout the paper 
labels a particular noise sequence $\xi_\alpha(t)$.
For any given $\alpha$,
there are infinitely many different trajectories $\vec q_\alpha(t)$ that
are distinguished by their initial conditions at some time $t_0$. Note that
no restriction on the dimensionality of the configuration space is imposed
and the derivations below work in one as well as in many dimensions. Although
in Eq.~\ref{LE}, the deterministic force is formally assumed to be derived
from a potential $U(\vec q)$, it would be straightforward to include
velocity-dependent forces from, e.g., a magnetic field.\cite{JFU00}
Meanwhile, the stochastic force $\vec\xi_\alpha(t)$ is assumed to be
Gaussian distributed with zero mean. 
It is related to the symmetric positive-definite friction matrix $\bm\Gamma$ 
through the fluctuation-dissipation theorem,
\cite{mori65,ford65,kubo66,fox78,zwan01}
\begin{equation}
  \label{FDTWhite}
  \left<\vec\xi_\alpha(t)\vec\xi_\alpha^\text{T}(t')\right>_\alpha = 
      2\kb T\,\bm{\Gamma} \, \delta(t-t') \;,
\end{equation}
where the angular brackets denote the average over the instances $\alpha$
of the fluctuating force, and the Dirac $\delta$ function appears
because white noise is local in time.

The assumption of white noise in Eq.~\ref{FDTWhite} is consistent
with the LE:
It requires that the dynamics of the heat bath, which determines the
correlation time of the fluctuating force, takes place on much shorter time
scales than the dynamics along the chosen system coordinates
---{\it viz.}, the reaction coordinate and any other 
solute or solvent modes coupled   to it.
This is not always
the case in applications relevant to chemistry where the dynamical
time scales of the system and the bath are usually comparable. 
Nevertheless, the special case of white 
noise is treated in detail in the next two sections
because it accommodates an explicit description of the
relevant geometric structures. 
The generalization of the moving TS structures to the GLE 
is formally straightforward though mathematically more cumbersome.
In the case of colored noise, the fluctuation dissipation theorem
requires that the friction kernel be nonlocal in time, {\it viz.}~contain
memory.\cite{mori65,ford65,kubo66} 
The TS trajectory is nevertheless computable and is presented in
section~\ref{sec:Coloured}. 
Perhaps surprisingly,
each of the geometric structures associated with the GLE is
analogous to that in the LE.

A further assumption in this work requires that the
the reactant and product regions in configuration space are
separated by a potential barrier, as is the case for most chemical
reactions. The position of the barrier is marked by a saddle point $\vec
q_0^\ddag=0$ of the potential $U(\vec q)$. In the absence of noise, the
saddle point is a fixed point of the dynamics.
The invariant manifolds that determine the
relevant phase space geometry \cite{bristol1,bristol2,Jaffe05} are attached
to it.
In the presence of a fluctuating force, the geometric structures
fixed at the saddle point no longer determine the reaction.
As shown below, however, there is
a unique stochastic trajectory $\vec q_\alpha^\ddag(t)$ that can play the role
of the saddle point and serve as the carrier of invariant manifolds.

Because the reaction rate is determined by the dynamics in a small
neighborhood of the saddle point,\cite{Truhlar96,Miller} 
the deterministic force in the LE of Eq.~\ref{LE} 
can be linearized around the saddle point to yield
\begin{equation}
  \label{linLE}
  \ddot{\vec q}_\alpha(t) = \bm{\Omega} \vec q_\alpha(t) -
        \bm{\Gamma}\dot{\vec q}_\alpha(t) + \vec\xi_\alpha(t)\;,
\end{equation}
where the first term in the RHS involves the 
symmetric force constant matrix,
\begin{equation}
  \bm{\Omega}_{ij}=-\left.\frac{\partial^2 U}{\partial q_i\partial q_j}\right|
      _{\vec q=\vec q^\ddag} \;.
\end{equation}
Although any rectilinear coordinate system can be chosen, it is convenient
to choose the one for which the force matrix,
\begin{equation}
  \bm{\Omega} = \begin{pmatrix}
              \omega_\text{b}^2 \\
              & -\omega_2^2 \\
              && \ddots \\
              &&& -\omega_N^2
           \end{pmatrix}\;,
\end{equation}
is diagonal from the outset.
In Eq.~\ref{linLE},
$q_1$ is the reaction coordinate,
$\omega_\text{b}$ the barrier frequency and $\omega_2, \dots\omega_N$ the
frequencies of oscillations in the stable transverse normal modes
$q_2, \dots q_N$.

\section{The fundamental construction}
\label{sec:fund}

The aim in the following is first to construct a 
no-recrossing surface 
for Eq.~\ref{linLE} and then to generalize the construction to
correlated noise. 
It is well known that in the absence of noise
and of dissipation
---which are closely related by Eq.~\ref{FDTWhite}---
the phase-space hypersurface ($q_1=0$) is free of 
recrossings.\cite{Miller,Truhlar96,bristol1,bristol2}.
As shown below, the
damping does not pose particular difficulties. The noise, however, does,
because the fluctuating force leads the particle to move randomly back and
forth, so that it will typically cross and recross any fixed dividing
surface many times. We will overcome this difficulty by
constructing a dividing surface that is itself randomly moving such as to
avoid recrossing. To achieve this, the dividing surface must be constructed
as a function of the fluctuating force.

Given any two
trajectories $\vec q_\alpha(t)$ and $\vec q_\alpha^\ddag(t)$ under the
influence of the same fluctuating force, we define the
relative coordinate,
\begin{equation}
  \label{relCoord}
  \Delta\vec q(t)=\vec q_\alpha(t)-\vec q_\alpha^\ddag(t) \;.
\end{equation}
It describes the location of the trajectory $\vec q_\alpha(t)$ with respect
to the moving origin $\vec q_\alpha^\ddag(t)$. By Eq.~\ref{linLE},
the relative coordinate $\Delta\vec q(t)$ satisfies the deterministic
equation of motion
\begin{equation}
  \label{relEq}
  \Delta\ddot{\vec q} 
      = \bm{\Omega}\,\Delta\vec q-\bm{\Gamma}\,\Delta\dot{\vec q}
\end{equation}
and thus exhibits non-random noiseless dynamics.  
This result is indicated by the lack of a subscript $\alpha$  
in $\Delta\vec q$.
Due to the
presence of the damping in Eq.~\ref{relEq}, the dynamics of the relative
coordinate, though noiseless, is still dissipative.
Nevertheless, 
with respect to the relative dynamics, 
it is possible to specify a no-recrossing surface,
as well as the invariant manifolds 
of the equilibrium point $\Delta\vec q=0$. 
These geometric objects in the noiseless phase space can
be regarded as being attached to and propagating with
the reference trajectory $\vec q_\alpha^\ddag(t)$.
They thus define
moving invariant manifolds and a randomly moving no-recrossing surface in
the phase space of the original, noisy system.

The relative dynamics between any two trajectories,
$\vec q_\alpha(t)$ and $\vec q_\alpha^\ddag(t)$, 
is described by Eq.~\ref{relEq}.
Because it is always noiseless, 
a moving no-recrossing surface can be thought of as attached to any
reference trajectory. 
However, only a specific surface is relevant for the
reaction dynamics: The crossing of the surface should indicate the
transition of the trajectory $\vec q_\alpha(t)$ from the reactant to the
product side of the barrier. It cannot serve that purpose if it is attached
to a reference trajectory that is itself far away from the barrier. 
If the
reference trajectory is chosen arbitrarily, it will typically descend into
either the reactant or the product wells over time and thus lose its
ability to carry a chemically meaningful dividing surface. 
Only a reference trajectory that
remains in the vicinity of the barrier for all time without ever descending
on either side of it can carry a no-recrossing surface that describes the
reaction in the same way that a static dividing surface in conventional TST
does.  Indeed, we will show below that for each instance of the noise there
is a unique reference trajectory with this property.  This is the
Transition State (TS) Trajectory mentioned in the introduction.  We will
henceforth restrict the notation $\vec q_\alpha^\ddag(t)$ to this
particular privileged reference trajectory.

The definition of the TS trajectory as ``remaining close to the saddle
point'' for all time might at first appear somewhat vague. 
In the construction of the TS trajectory below, 
it will become clear that the
exponential instability that is inherent in the noiseless dynamics
associated with Eq.~\ref{linLE} in both the distant past and the remote
future must be absent from the TS trajectory.
A precise
meaning can be given to the notion in statistical terms as follows: Because
it is stochastic, at any given time there is a probability distribution for the
TS trajectory in phase space. This distribution is invariant under the time
evolution given by Eq.~\ref{linLE}.
In mathematical language, the TS trajectory represents an invariant measure
of the noisy dynamical system.\cite{Arnold98} 
That is, for any instance $\vec\xi_\alpha(\cdot)$ of the
noise, where the dot indicates a function of time, there is an instance
$\vec q_\alpha^\ddag\left[\vec\xi_\alpha(\cdot)\right](t)$ of the TS
trajectory that is given as a functional of the noise and is itself a
function of time, $t$. 
It can be used to obtain the instance of the TS
trajectory $\vec q_\alpha^\ddag\left[\vec\xi_\alpha(\cdot)\right](t+\tau)$ that
will be found if the origin 
from which time is measured is moved an increment $\tau\ne 0$. 
Alternatively, one can use the time-shifted noise
$\vec\xi_\alpha(\cdot+\tau)$ and compute the corresponding TS trajectory 
$\vec q_\alpha^\ddag\left[\vec\xi_\alpha(\cdot+\tau)\right](t)$. 
The TS trajectory is
uniquely characterized by the requirement that these two ways of computing
the time shift agree: 
$\vec q_\alpha^\ddag\left[\vec\xi_\alpha(\cdot)\right](t+\tau) 
  = \vec q_\alpha^\ddag\left[\vec\xi_\alpha(\cdot+\tau)\right](t)$.

We have now given a general outline of the method. In the following
sections, we will actually construct the TS trajectory, determine its
statistical properties and describe the invariant manifolds and the
no-recrossing surface attached to it in detail.

\section{The Transition State in white noise}
\label{sec:WN}

Equation \ref{linLE} can be rewritten as a first-order
equation of motion in the $2N$-dimensional phase space with the coordinates
\begin{equation}
  \label{xDef}
  \vec z=\begin{pmatrix}
            \vec q \\
            \vec v
         \end{pmatrix} \;,
\end{equation}
where $\vec v=\dot{\vec q}$, as
\begin{equation}
  \label{phaseLE}
  \dot{\vec z}_\alpha(t)=A \vec z_\alpha(t)+
                        \begin{pmatrix}
                           0 \\
                           \vec\xi_\alpha(t)
                        \end{pmatrix}
\end{equation}
with the $2N$-dimensional constant matrix
\begin{equation}
  \label{ADef}
  A=\begin{pmatrix}
          \bm{0} & \bm{I}\\
          \bm{\Omega} & -\bm{\Gamma}
    \end{pmatrix}\;,
\end{equation}
where $\bm{I}$ is the $N\times N$ identity matrix.
After diagonalizing the matrix $A$, its eigenvalues 
are denoted by $\epsilon_j$ and its
corresponding eigenvectors by $\vec V_j$. The Langevin
equation, Eq.~\ref{phaseLE}, then decomposes into a set of $2N$ independent
scalar equations of motion
\begin{equation}
  \label{diagEq}
  \dot z_{\alpha j}(t) = \epsilon_j z_{\alpha j}(t) + \xi_{\alpha j}(t) \;,
\end{equation}
where $z_{\alpha j}$ are the components of $\vec z$ in the basis $\vec V_j$
of eigenvectors of $A$ and $\xi_{\alpha j}$ the corresponding components of
$(0,\vec\xi_\alpha(t))^{\text{T}}$.

\subsection{Relative dynamics}
\label{ssec:WNnoiseless}

The noiseless equation of motion for the relative coordinate $\Delta\vec
q(t)$ in
Eq.~\ref{relCoord}, when lifted into phase space via
Eq.~\ref{xDef}
and expressed in diagonal coordinates, reads
\begin{equation}
  \label{DeltaXEq}
  \Delta\dot z_j(t) = \epsilon_j \Delta z_j(t) \;.
\end{equation}
It is solved by
\begin{equation}
  \label{homSolComp}
  \Delta z_j(t) = e^{\epsilon_j t} \Delta z_j(0)
\end{equation}
with a set of integration constants $\Delta
z_j(0)$. Equation~\ref{homSolComp} leads to a solution of the form
\begin{equation}
  \label{homSolWhite}
  \Delta\vec q(t) = \sum_j c_j \vec v_j e^{\epsilon_j t}
\end{equation}
for the relative coordinate vector $\Delta\vec q(t)$, where $\vec v_j$ are
the configuration space parts of the phase space eigenvectors $\vec V_j$
and the $c_j$ are arbitrary constants.
To identify reactive trajectories, one must consider the
behavior of $\Delta\vec q(t)$ in the remote future and in the distant past
and then determine those trajectories that pass across the barrier from 
the reactant to the product side.

\begin{figure*}
  \centerline{\includegraphics[width=.3\textwidth]{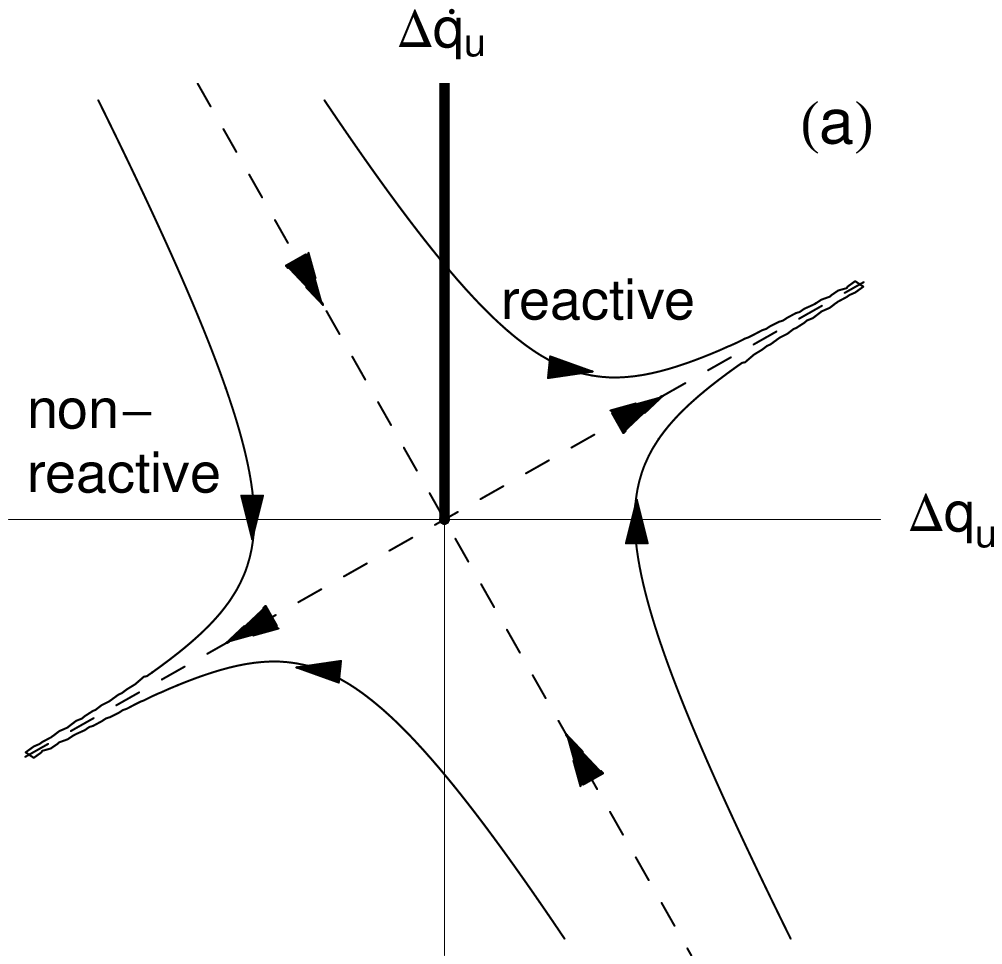}\hfill
              \includegraphics[width=.3\textwidth]{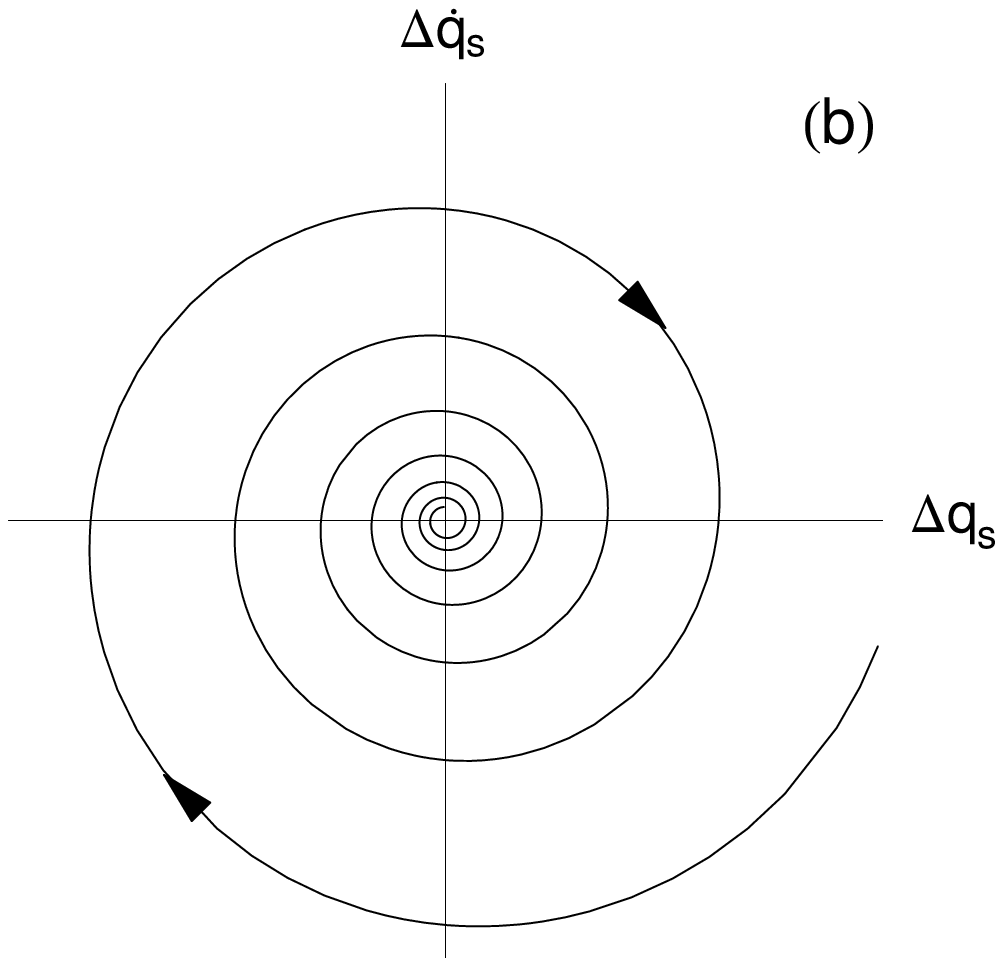}\hfill
              \includegraphics[width=.3\textwidth]{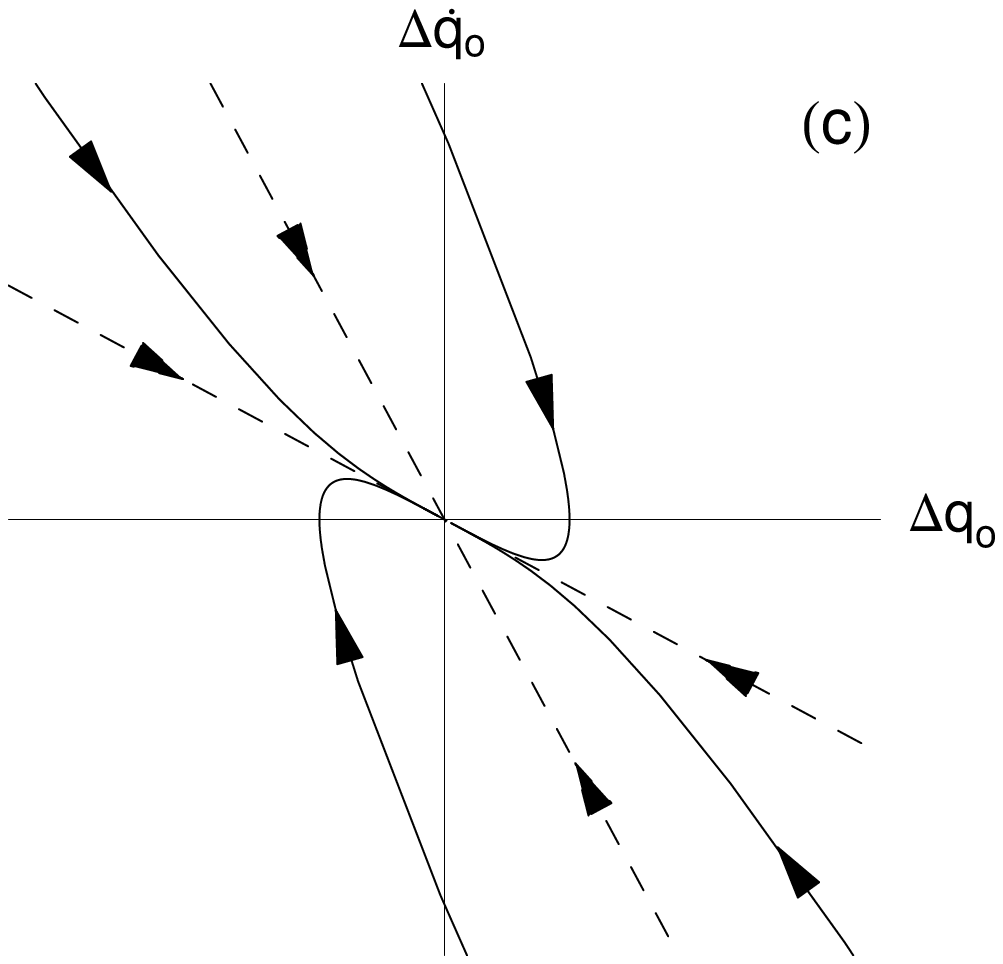}}
  \caption{Phase portrait of the noiseless damped dynamics
     corresponding to the linear Langevin equation~(\ref{linLE}) in
     (a) the unstable reactive degree of freedom, (b) a transverse stable
     degree of freedom, (c) an overdamped transverse degree of freedom.}
  \label{fig:phase}%
\end{figure*}

The simplest case occurs in one degree of freedom where
$\bf\Omega$
reduces to the squared barrier frequency 
$\omega_\text{b}^2$ 
and
$\bm\Gamma$
to a scalar damping coefficient 
$\gamma$.
In this case, the matrix $A$ can be diagonalized analytically. 
It possesses the eigenvalues
\begin{subequations}
\label{EVals}
\begin{align}
  \label{epsu}
  \epsilon_\text{u}&=
      -\frac 12\left(\gamma-\sqrt{\gamma^2+4\omega_\text{b}^2}\right) > 0 \\
\intertext{and}
  \label{epss}
  \epsilon_\text{s}&=
      -\frac 12\left(\gamma+\sqrt{\gamma^2+4\omega_\text{b}^2}\right) < 0
\end{align}
\end{subequations}
with the corresponding eigenvectors
\begin{equation}
  \label{EVecs}
  \vec V_\text{u}=\begin{pmatrix} 1 \\ \epsilon_\text{u} \end{pmatrix}
  \qquad \text{and} \qquad
  \vec V_\text{s}=\begin{pmatrix} 1 \\ \epsilon_\text{s} \end{pmatrix} \;.
\end{equation}
The phase portrait of the dynamics is shown in
Fig.~\ref{fig:phase}a. The eigenvectors corresponding to the positive
and negative eigenvalues, respectively, span one-dimensional unstable and
stable manifolds of the saddle point. They act as separatrices between
reactive and nonreactive trajectories. The knowledge of the invariant
manifolds allows one to determine the ultimate fate of a specific
trajectory from its initial conditions. 
It should be clear that line, $\Delta q_\text{u}=0$,
in the quadrant of reactive trajectories acts as a surface of no
recrossing. 

In the absence of damping (and in units where $\omega_\text{b}=1$), the
invariant manifolds bisect the angles between the coordinate axes. The
presence of damping destroys this symmetry. 
As the damping constant $\gamma$ increases, 
on obtains the limits, $\epsilon_\text{s}\to 0$ and
$\epsilon_\text{u}\to\infty$. 
Therefore, the unstable manifold rotates
toward the $q_\text{u}$-axis, the stable manifold toward the 
$\dot q_\text{u}$-axis. 
In the limit of infinite damping, 
the invariant manifolds coincide with the axis. 
Thus, the fraction of phase space on the reactant side that
corresponds to reactive trajectories decreases with increasing
friction. This is intuitively clear because if the dissipation is strong, a
trajectory must start at a given initial position with a high velocity to
overcome the friction and cross the barrier.

In multiple dimensions, the dynamics of the single unstable degree of
freedom is the same as in the one-dimensional case shown in
Fig.~\ref{fig:phase}a. Transverse damped oscillations, as shown in
Fig.~\ref{fig:phase}b, must be added to this picture. Their presence
manifests itself, for small damping, through $N-1$ complex conjugate pairs
of eigenvalues $\epsilon_j$. For stronger damping, some of the transverse
modes can become overdamped, as illustrated in figure~\ref{fig:phase}c,
so that further eigenvalues become negative real. In any case, there is
exactly one unstable eigenvalue with positive real part. This eigenvalue is
actually real and corresponds to the system sliding down the barrier. 
In all other directions, the dynamics is stable, and at least one eigenvalue
is negative real.  The dynamics in the distant past is determined by the
eigenvalue or pair of eigenvalues with the largest
negative real part.

The eigenvector corresponding to the most stable eigenvalue together with
the unstable eigenvector span a plane in phase space in which the dynamics
is given by the phase portrait of Fig.~\ref{fig:phase}a. 
Because the dynamics in all other, ``transverse'' directions is stable, the
separatrices between reactive and nonreactive trajectories that we
identified for the one-dimensional dynamics together with the transverse
subspace form separatrices in the high-dimensional phase space. In a
similar manner, a no-recrossing curve in the plane together with the
transverse directions form a no-recrossing surface in the full phase space.

If the most stable eigenvalues occur as a complex conjugate pair, the
leading dynamics in the remote past is given by an oscillation across the
barrier that, because it is strongly damped, has a huge amplitude in the
distant past and causes the system to cross the barrier infinitely
often. This behavior is clearly unphysical. In fact, once the oscillations
become large, nonlinearities of the reaction dynamics become relevant, and
the ultimate fate of the trajectory can no longer be determined from the
linear approximation. 
This situation can only occur if
the damping is strong and
at the same time the undamped transverse frequencies are so large that the
corresponding eigenmodes do not become overdamped. 
This rare situation is ignored below.

An exception to the discussion above occurs if for any eigenvector $\vec
v_j$ the component $v_j^{(1)}$ in the direction of the reaction coordinate
is zero. In this case the excitation of the corresponding eigenmode has no
impact on whether a trajectory is reactive or nonreactive. The unstable and
the most stable eigenvalues must then be chosen among the eigenvalues with
$v_j^{(1)}\ne 0$, which we call the ``relevant'' eigenvalues.  Because the
unstable eigenmode corresponds to the particle sliding down the barrier, it
is always relevant. Similarly, there must be one relevant negative
eigenvalue corresponding to the particle being shot at the barrier from
infinity and coming to rest on it.  The occurrence of irrelevant
eigenvalues might seem a rather unlikely exception, but it does arise in
the important special case of isotropic friction, where the friction
coefficient $\bm{\Gamma}=\gamma\bm{I}$ is a scalar multiple of the identity
matrix. In this situation, the friction does not couple different normal
modes of the undamped dynamics, and all but two eigenvalues are
irrelevant. In particular, our construction of a no-recrossing surface can
always be carried out for isotropic friction.

\subsection{The Transition State Trajectory}
\label{ssec:WNref}

The influence of noise in the equation of motion,
Eq.~\ref{diagEq}, can be
accounted for by means of the retarded Green's function
\begin{equation}
  \label{Gret}
  G^\text{ret}(t) = \begin{cases}
                       0 & \text{if } t<0 \\
                       e^{\epsilon_j t} & \text{if } t>0 \;,
                    \end{cases}
\end{equation}
which satisfies
\begin{equation}
  \label{Geq}
  \dot G^\text{ret}(t) - \epsilon_j G^\text{ret}(t) = \delta(t) \;.
\end{equation}
It gives the general solution of Eq.~\ref{diagEq} as
\begin{align}
  \label{retSol}
    z_{\alpha j}(t) &= c_{\alpha j} e^{\epsilon_j t} +
      \int_{-\infty}^\infty G^\text{ret}(t-\tau) \xi_{\alpha j}(\tau)\,d\tau
      \nonumber\\
     &= c_{\alpha j} e^{\epsilon_j t} +
       \int_{-\infty}^t e^{\epsilon_j(t-\tau)} \xi_{\alpha j}(\tau)\,d\tau
\end{align}
with arbitrary constants $c_{\alpha j}$. The integral
\begin{align}
  \label{xiStab}
    z^\ddag_{\alpha j}(t)
        &= \int_{-\infty}^t e^{\epsilon_j(t-\tau)}\xi_{\alpha j}(\tau)\,d\tau
          \nonumber\\
        &= \int_{-\infty}^0 e^{-\epsilon_j \tau} \xi_{\alpha j}(t+\tau)\,d\tau
\end{align}
exists with probability one
if $\Re\epsilon_j<0$, i.e., for the stable components. Only for
those, therefore, is the solution, Eq.~\ref{retSol}, meaningful.

The first term in Eq.~\ref{retSol} tends to zero as $t\to\infty$,
but it diverges exponentially as $t\to-\infty$. The only solution to
Eq.~\ref{diagEq} that remains bounded in the remote past is obtained
by setting $c_{\alpha j}=0$. 
But the TS Trajectory has been defined as the
unique trajectory that remains in the neighborhood of the saddle point at
all times. It now becomes clear that this condition indeed determines the
stable components of the TS trajectory uniquely and that they are given
by Eq.~\ref{xiStab}. Because they depend on the noise $\xi_{\alpha j}(\cdot)$,
the $z^\ddag_{\alpha j}(t)$ are stochastic functions of time. From the second
form given in Eq.~\ref{xiStab}, it should be obvious that the probability
distribution of $z^\ddag_{\alpha j}$ is stationary in time if the
distribution of $\vec\xi_\alpha(\cdot)$ is itself stationary.

The solution in Eq.~\ref{retSol} fails for the unstable component of the
trajectory, where $\Re\epsilon_j>0$. To compute that component, we use the
advanced Green's function
\begin{equation}
  \label{Gadv}
  G^\text{adv}(t) = \begin{cases}
                      -e^{\epsilon_j t} & \text{if } t<0 \\
                      0 & \text{if } t>0
                    \end{cases}
\end{equation}
to obtain the solution
\begin{equation}
  \label{advSol}
  z_j(t) = c_{\alpha j} e^{\epsilon_j t} + z^\ddag_{\alpha j}(t)
\end{equation}
with constants $c_{\alpha j}$ and
\begin{align}
  \label{xiUnst}
    z^\ddag_{\alpha j}(t) 
      &= -\int_t^\infty e^{\epsilon_j(t-\tau)} \xi_{\alpha j}(\tau)\,d\tau
         \nonumber\\
      &= -\int_0^\infty e^{-\epsilon_j\tau} \xi_{\alpha j}(t+\tau) \, d\tau \;.
\end{align}
The position in Eq.~\ref{advSol} is bounded in the remote future only if
$c_{\alpha j}=0$. Therefore, $z^\ddag_{\alpha j}(t)$ given in
Eq.~\ref{xiUnst} can be identified as the component of the TS
Trajectory in the unstable direction. Notice that the stable
components in Eq.~\ref{xiStab} at time $t$ depend on the past 
of the fluctuating force, whereas the 
unstable component in Eq.~\ref{xiUnst} depends on its future.

Through Eqs.~\ref{xiStab} and \ref{xiUnst}, all components of the TS
trajectory $\vec z_\alpha^\ddag(t)$ in phase space have been specified. The
TS trajectory $\vec q_\alpha^\ddag(t)$ in configuration space can be
obtained because the transformation connecting the
position-velocity coordinates and diagonal coordinates is known. The
consistency of the calculation can then be checked by verifying that the
velocity obtained from the coordinate transformation is actually the time
derivative of the position coordinate.

\begin{figure*}
  \centerline{\includegraphics[width=.48\textwidth]{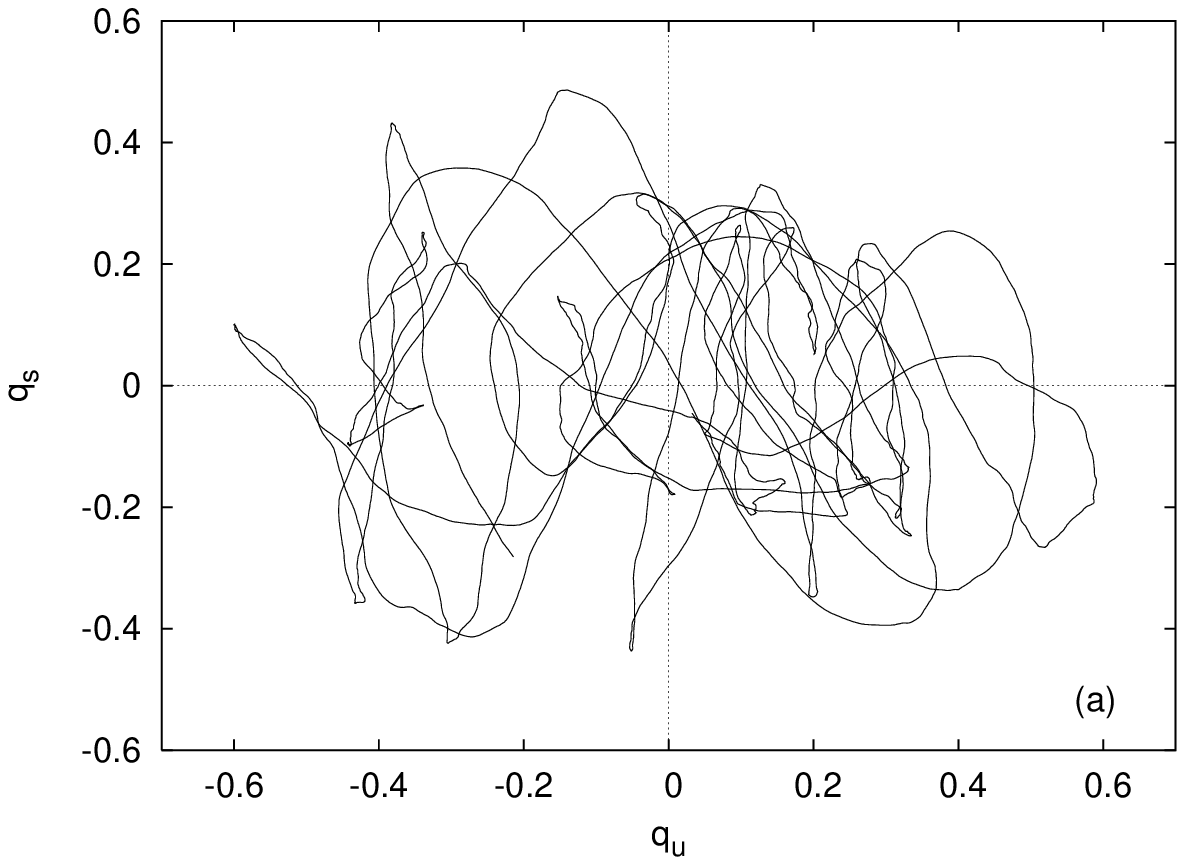}\hfill
              \includegraphics[width=.48\textwidth]{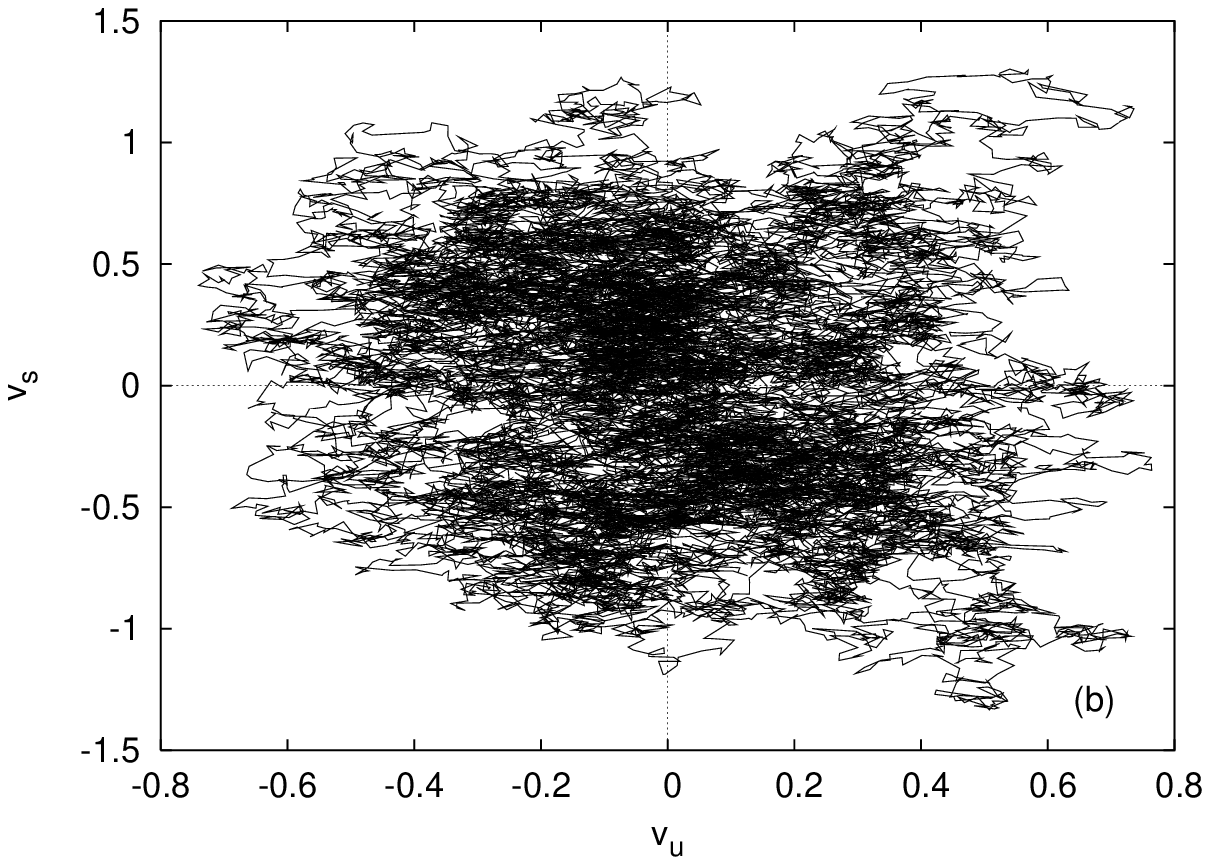}}
  \centerline{\includegraphics[width=.48\textwidth]{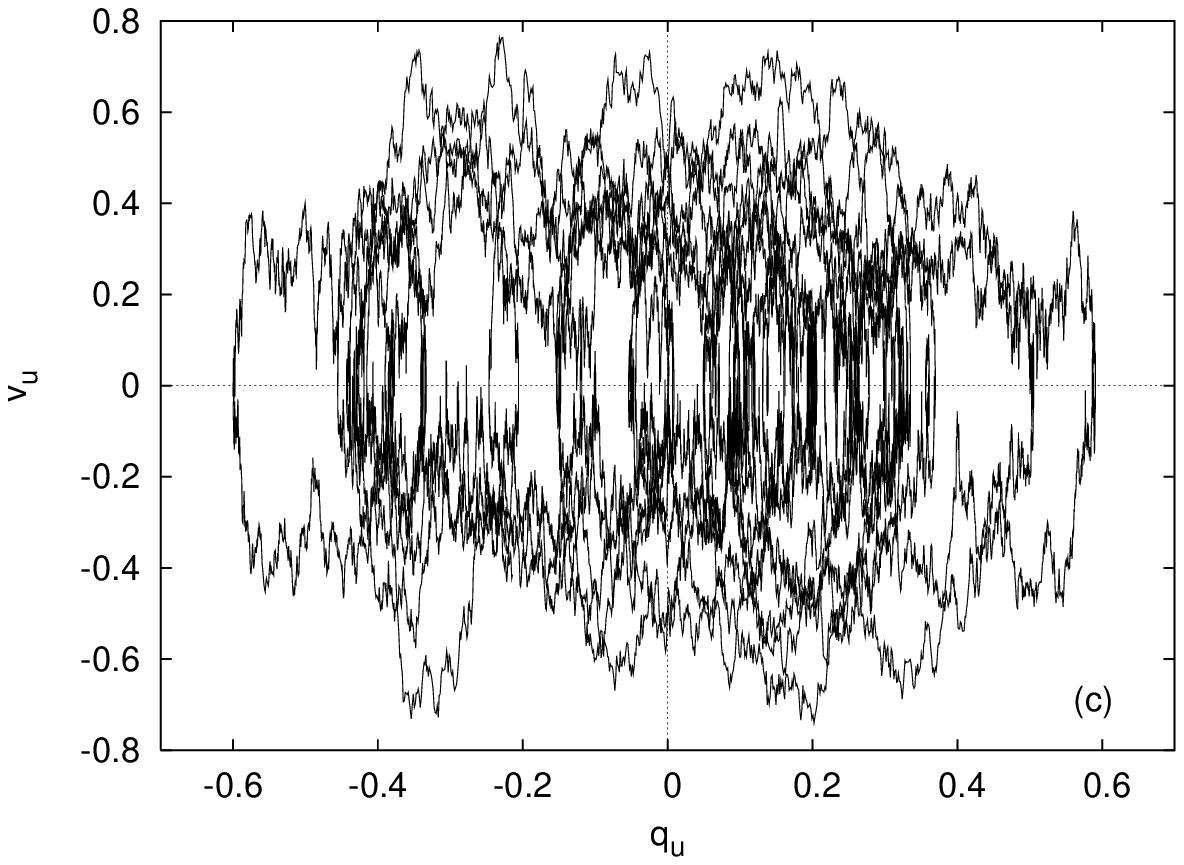}\hfill
              \includegraphics[width=.48\textwidth]{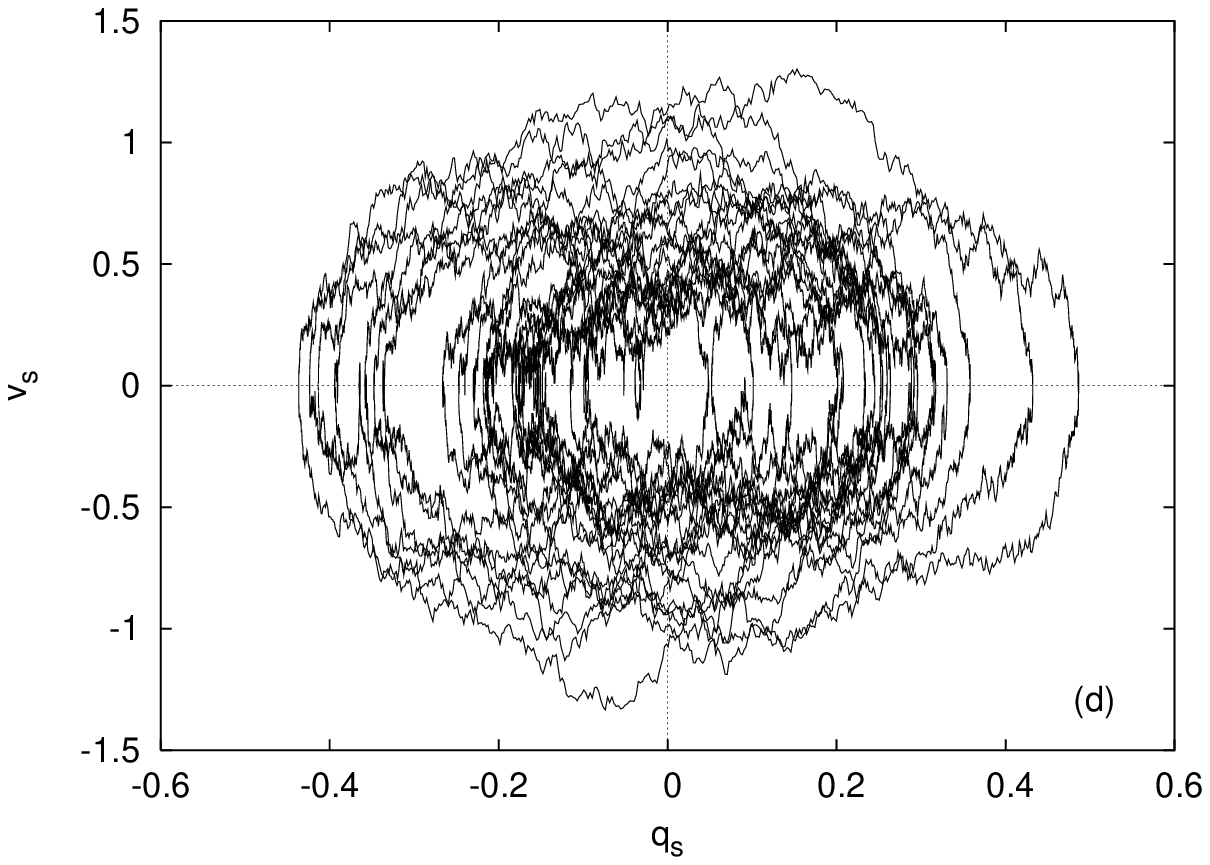}}
  \caption{\label{fig:TSProj}
    A random instance of the TS trajectory in a system with $N=2$ degrees
    of freedom, projected onto (a) configuration space, (b) velocity space,
    (c) the reactive degree of freedom, and (d) the transverse degree of
    freedom.  Units have been chosen so that $\omega_\text{b}=1$, $\kb
    T=1$. The transverse frequency is $\omega_2=2$, and friction is
    isotropic, $\bm{\Gamma}=\gamma\bm{ I}$ with $\gamma=0.5$.
  }
\end{figure*}

A specific instance of the TS trajectory for a system with two degrees of
freedom (and a four-dimensional phase space) is shown in
Fig.~\ref{fig:TSProj}. It can be observed how the trajectory randomly
wanders around in the vicinity of the saddle point without ever leaving
it. The velocity space projection of the TS trajectory appears much more
noisy than its configuration space projection. 
This results because 
the environment is assumed to cause a fluctuating force 
in Eq.~(\ref{LE})
that couples directly only to the velocity, but not to the
position. 
It should be clear from the above discussion, however, that this
condition can easily be relaxed.
In the absence of noise and of damping, 
any trajectory would exhibit an elliptical orbit in the projection
onto the stable degree of freedom
This behavior is still
visible in the noisy situation of figure~\ref{fig:TSProj}d where a slow
change of energy, corresponding to the size of the ellipse, is superimposed
over the elliptic motion.

\begin{figure*}
  \centerline{\includegraphics[width=.3\textwidth]{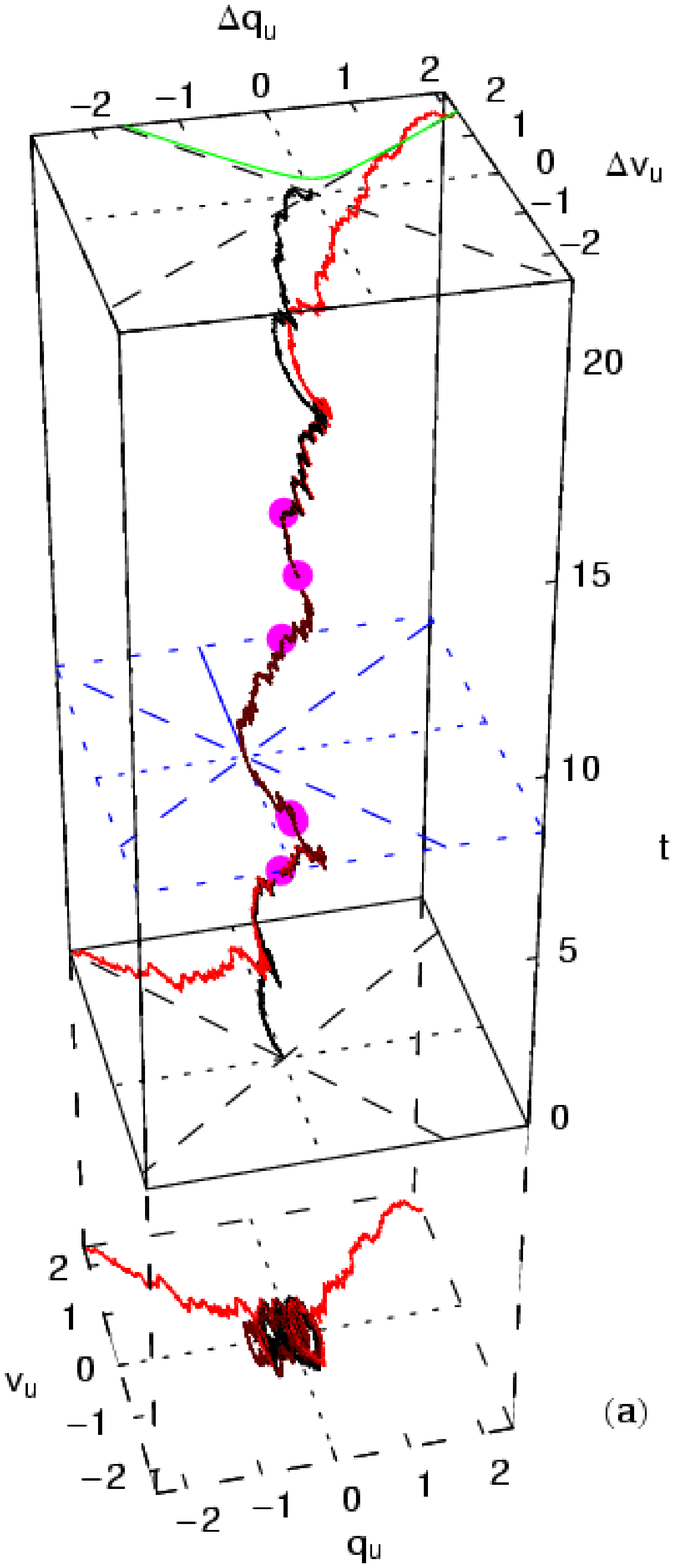}\hfill
              \includegraphics[width=.3\textwidth]{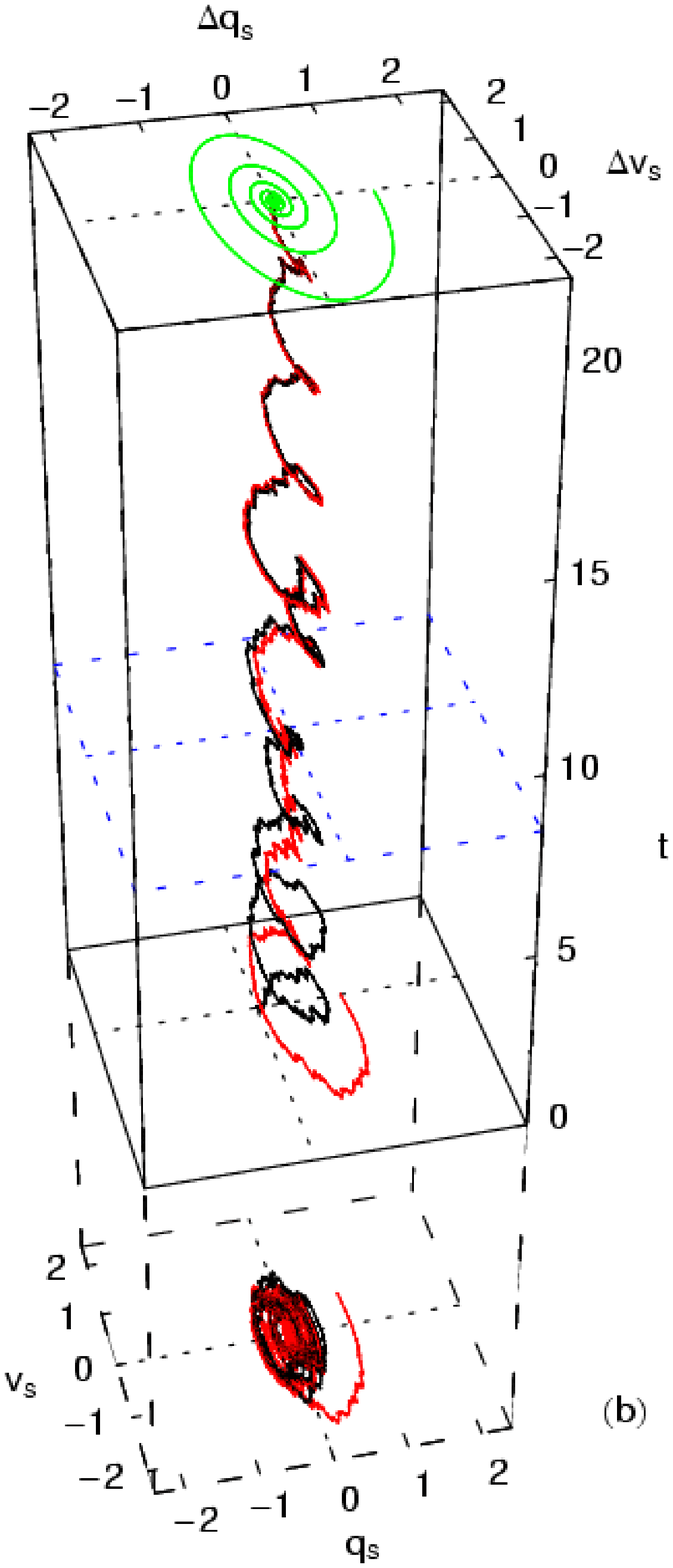}\hfill
              \includegraphics[width=.3\textwidth]{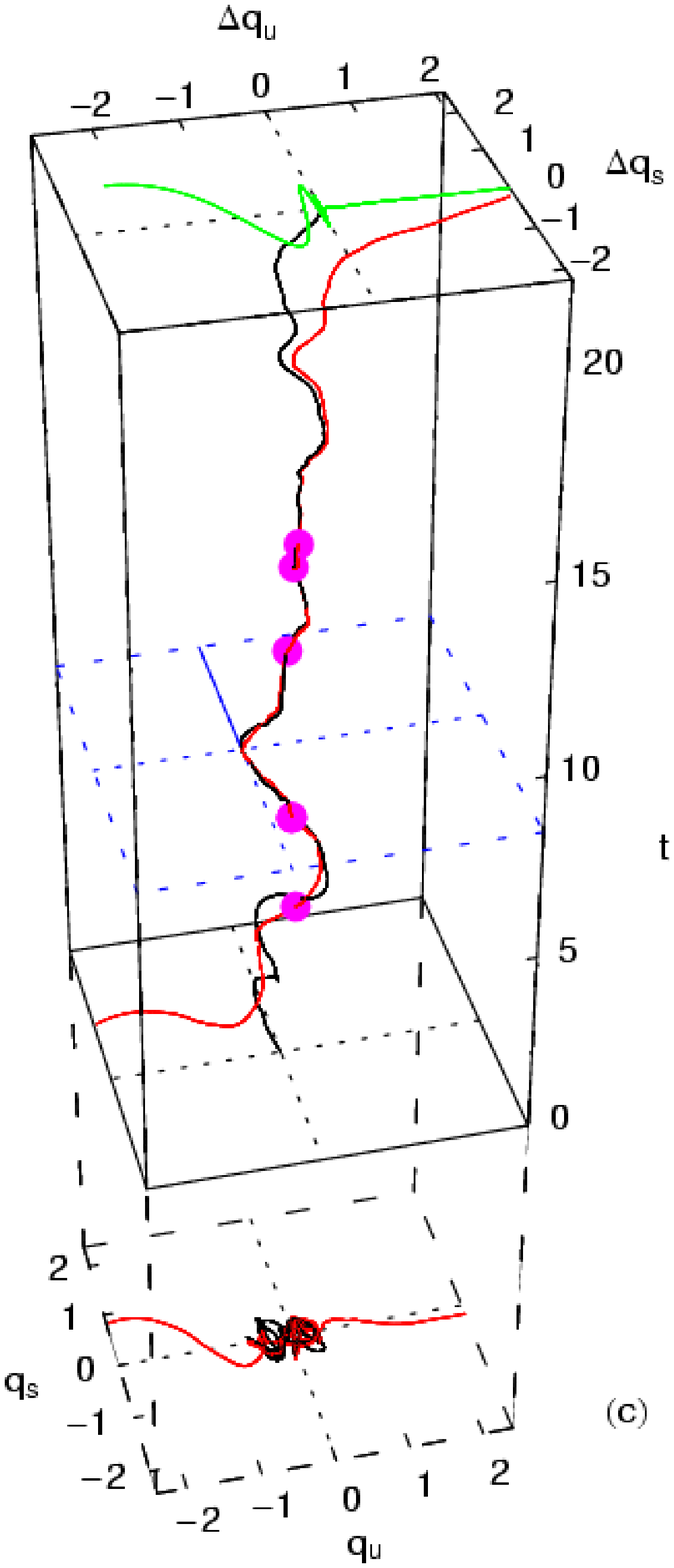}}
  \caption{\label{fig:Traj2D}
    The instance of the TS trajectory shown in Fig.~\ref{fig:TSProj} and
    a transition path under the influence of the same noise. Trajectories
    are projected onto (a) the reactive degree of freedom, (b) the
    transverse degree of freedom, and (c) configuration space. At the lower
    end of each column is a projection into the corresponding phase-space
    plane, disregarding time. The instantaneous positions of the moving
    coordinate axes (dotted) and, in (a), of the invariant manifolds
    (dashed) are shown at $t=0$, at $t=22$, and at the unique reaction time
    $t_\text{react}=8.500$ where the TS is crossed. The solid line at
    $t_\text{react}$ indicates the moving TS The curves in the top faces
    of each panel illustrates the dynamics of the transition path in
    relative coordinates $\Delta q$, $\Delta v$ (not to scale in (a), for
    graphical reasons).
  }
\end{figure*}

As described in section~III, the TS trajectory serves as the
origin of a moving coordinate system, and the geometric structures in the
noiseless phase space that we have described in
section~\ref{ssec:WNnoiseless} are attached to it. This is illustrated in
Fig.~\ref{fig:Traj2D} for two degrees of freedom. This case already
exhibits all salient features of the dynamics in arbitrary high
dimensions. The figure compares a specific instance of the TS trajectory
with a transition path under the influence of the same noise. It can be
seen how the transition path approaches the TS trajectory from the reactant
side, remains in its vicinity for a while and then wanders off to the
product side. Because the moving invariant manifolds of the TS trajectory
are known, it can be predicted already at time $t=0$ that the trajectory
actually will lead to a reaction instead of returning to the reagent side
of the saddle.  From both the time-dependent plots and the time-independent
projections in Fig.~\ref{fig:Traj2D}a and \ref{fig:Traj2D}c 
it is clear that the
transition path crosses the space-fixed TS surface
$q_\text{u}=q_\text{u}^\ddag=0$ several times before it finally descends on
the product side. The moving TS surface $\Delta q_\text{u}=0$, by contrast,
is crossed only once, at the reaction time $t_\text{react}=8.500$. That
this is actually the case can be seen from the curves in the top
faces of each column, which indicate the noiseless relative motion $\Delta
q(t)$, $\Delta v(t)$ between the TS trajectory and the transition path. In
the relative coordinates we find, as expected, the hyperbolic motion known
from Fig.~\ref{fig:phase} in the unstable degree of freedom, a damped
oscillation in the stable transverse degree of freedom, and a superposition
of the two in the configuration space projection in
Fig.~\ref{fig:Traj2D}c.

\subsection{Statistical properties of the TS trajectory}
\label{ssec:WNstat}

If the noise $\vec\xi_\alpha(t)$ is assumed to be Gaussian with zero
mean, it is clear from Eqs.~\ref{xiStab} and \ref{xiUnst} that the
same is true for all components $z^\ddag_{\alpha j}$ of the TS trajectory
and, by extension, also for its position and velocity coordinates. To
specify their joint probability distribution completely, it thus remains to
calculate their time autocorrelation and cross-correlation functions.

The most general fluctuating force $\vec\xi_\alpha(t)$ satisfying the
fluctuation-dissipation theorem, Eq.~\ref{FDTWhite}, can be written as
\begin{equation}
  \label{WNTerm}
  \vec\xi_\alpha(t) = \bm{g} \vec W_\alpha(t)
\end{equation}
in terms of $M$ realizations of standard white noise $\vec
W_\alpha(t)=(W_{\alpha 1}(t), \dots, W_{\alpha M}(t))^\text{T}$ normalized to
\begin{equation}
  \label{WNCorrel}
     \left< W_{\alpha i}(t) W_{\alpha j}(s) \right>_\alpha
   = \delta(t-s)\,\delta_{ij}
\end{equation}
and an $N\times M$ matrix $g$ of coupling strengths satisfying
\begin{equation}
  \label{genWNCorr}
  \bm{g} \bm{g}^\text{T} = 2\kb T\,\bm{\Gamma} \;.
\end{equation}
The statistical properties of the noise are completely specified
by Eq.~\ref{genWNCorr}. If $M>N$, a representation can be found that has
fewer independent noise sources, but still satisfies Eq.~\ref{genWNCorr} 
with the same matrix $\bm\Gamma$. For that reason, it can be assumed without
loss of generality that $M\le N$.  The components of the noise in diagonal
coordinates can be written as
\begin{equation}
  \label{RjWN}
  \xi_{\alpha j}(t) = \sum_i \kappa_{ji} W_{\alpha i}(t)
\end{equation}
with a set of constants $\kappa_{ji}$ computed from the matrix $\bm{g}$ and
the coordinate transform between position-velocity coordinates and diagonal
coordinates.

With the noise term, Eq.~\ref{RjWN}, the time correlation function of two
stable components~(\ref{xiStab}) of the TS trajectory, with
$\Re\epsilon_j<0$, $\Re\epsilon_k<0$, reads
\begin{multline}
   \left<z^\ddag_{\alpha j}(t)z^\ddag_{\alpha k}(0)\right>_\alpha =
    \sum_{mn} \kappa_{jm}\kappa_{kn}
       \int_{-\infty}^t d\tau\, e^{\epsilon_j(t-\tau)}\,\times \\
       \int_{-\infty}^0 d\sigma\,e^{-\epsilon_k \sigma}
       \left<W_{\alpha m}(\tau) W_{\alpha n}(\sigma)\right>_\alpha \;.
\end{multline}
Using Eq.~\ref{WNCorrel}, for $t\ge0$ and with
\begin{equation}
  \label{KappaDef}
  K_{jk}=\sum_{n} \kappa_{jn}\kappa_{kn} \;,
\end{equation}
this can be evaluated to
\begin{subequations}
\label{WNCorr}
\begin{align}
   \left<z^\ddag_{\alpha j}(t)z^\ddag_{\alpha k}(0)\right>_\alpha &=
      \frac{K_{jk}}{-(\epsilon_j+\epsilon_k)}\,e^{\epsilon_j t}
      \hspace{-96pt}\nonumber \\
      &&\text{if }\Re\epsilon_j<0\text{, }\Re\epsilon_k<0.
      \label{WNStabCorr} \\
\intertext{Similarly,}
   \left<z^\ddag_{\alpha j}(t)z^\ddag_{\alpha k}(0)\right>_\alpha &=
      \frac{K_{jk}}{\epsilon_j+\epsilon_k}\,e^{\epsilon_k t} 
      \hspace{-96pt}\nonumber\\
      &&\text{if }\Re\epsilon_j>0\text{, }\Re\epsilon_k>0,
      \label{WNUnstCorr} \\
   \left<z^\ddag_{\alpha j}(t)z^\ddag_{\alpha k}(0)\right>_\alpha &=
      \frac{K_{jk}}{\epsilon_j+\epsilon_k}
          \left(e^{-\epsilon_k t}-e^{\epsilon_j t}\right)
	  \hspace{-96pt}\nonumber\\
      &&\text{if }\Re\epsilon_j<0\text{, }\Re\epsilon_k>0,
      \label{WNSUCorr} \\
\intertext{and}
   \left<z^\ddag_{\alpha j}(t)z^\ddag_{\alpha k}(0)\right>_\alpha &= 0
      &\text{if }\Re\epsilon_j>0\text{, }\Re\epsilon_k<0.
      \label{WNUSCorr}
\end{align}
\end{subequations}

All these correlation functions decay exponentially for long times. 
Note that that the correlation in Eq.~\ref{WNUSCorr} 
of an unstable component at a later
time $t>0$ with a stable component at time zero vanishes identically
because the former depends on the fluctuating force after time $t$, the
latter on the force before time zero. For the same reason, the
correlation in Eq.~\ref{WNSUCorr} of a stable component at $t>0$ with an
unstable component at time zero vanishes as $t\to 0$.

The correlation functions in Eq.~\ref{WNCorr} can be evaluated explicitly for
one-dimensional dynamics. In this case, the eigenvalues are given by
Eq.~\ref{EVals}, and the matrix $A$ can be diagonalized by
\begin{equation}
  \label{ADiag}
  A=M \begin{pmatrix} \epsilon_\text{s} & 0 \\ 0 & \epsilon_\text{u}
      \end{pmatrix}
    M^{-1}
\end{equation}
with
\begin{equation}
  \label{MDiag}
  M=\begin{pmatrix} 1 & 1 \\ \epsilon_\text{s} & \epsilon_\text{u}
    \end{pmatrix}
  \quad \text{and} \quad
  M^{-1} = \frac{1}{\epsilon_\text{u}-\epsilon_\text{s}} 
    \begin{pmatrix} \epsilon_\text{u} & -1 \\ -\epsilon_\text{s} & 1
    \end{pmatrix} \;.
\end{equation}
In one dimension, the fluctuating force in Eq.~\ref{WNTerm} can have only one
independent component with a scalar coupling strength $g$, and
using Eq.~\ref{MDiag} the diagonal coupling strength reduces to
\begin{equation}
  \label{kappaDef}
  \kappa := \kappa_\text{u}=-\kappa_\text{s} = 
            \frac{g}{\epsilon_\text{u}-\epsilon_\text{s}} \;.
\end{equation}

\begin{figure}
  \centerline{\includegraphics[width=.45\textwidth]{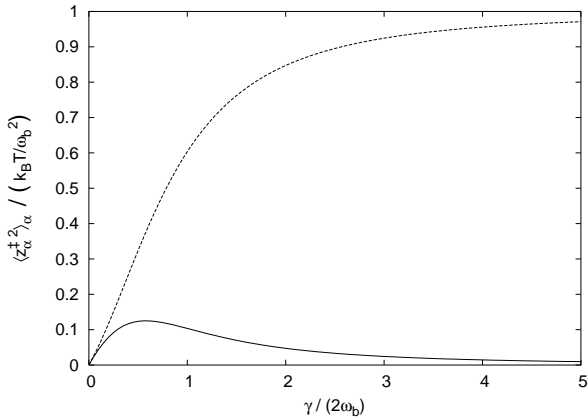}}
  \caption{\label{fig:SUVars}
     Variances~(\ref{SUVars}) of stable (solid curve) and unstable (dashed
     curve) components the TS trajectory $z_\alpha^\ddag(t)$ in one dimension.
  }
\end{figure}

We will in the following only discuss the correlation
functions in Eq.~\ref{WNCorr} for vanishing time shift $t=0$. 
The covariances of the stable and unstable components of one-dimensional
dynamics reduce to
\begin{subequations}
\label{SUVars}
\begin{align}
  \left<z^{\ddag 2}_\text{u}\right>_\alpha 
        &= \frac{\kappa^2}{2\epsilon_\text{u}} \;, \label{UVar}\\
  \left<z^{\ddag 2}_\text{s}\right>_\alpha
        &= \frac{\kappa^2}{-2\epsilon_\text{s}} \;, \label{SVar}\\
  \left<z^\ddag_\text{u}z^\ddag_\text{s}\right>_\alpha &= 0 \;.
\end{align}
\end{subequations}
The variances, Eq.~\ref{SUVars}, of the stable and unstable components are
shown in Fig.~\ref{fig:SUVars} as a function of the damping constant
$\gamma$, which is proportional to the noise strength. Not surprisingly,
both variances vanish in the absence of noise $(\gamma=0)$. In the limit of
strong noise, the variance of the stable component tends to zero as
$1/\gamma^2$, whereas the variance of the unstable component approaches its
maximum value $\kb T/\omega_\text{b}^2$.

\begin{figure}
  \centerline{\includegraphics[width=.45\textwidth]{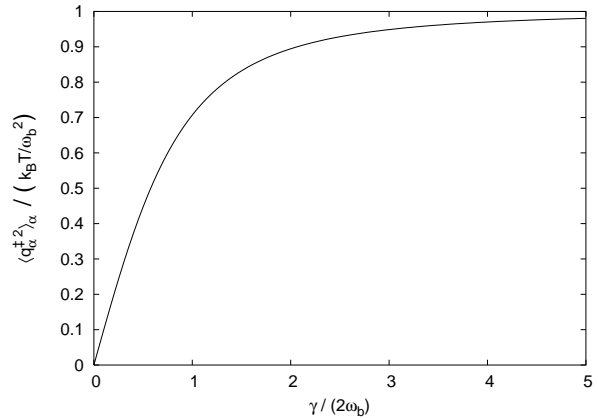}}
  \caption{\label{fig:QVars}
     Variance~(\ref{SigmaDef}) of the position
     coordinate of the TS trajectory $z_\alpha^\ddag(t)$ in one dimension.
  }
\end{figure}

Using the coordinate transformation in Eq.~\ref{MDiag}, the diagonal
covariances in Eq.~\ref{SUVars} can be converted into the covariances of the
position $q_\alpha^\ddag$ and velocity $v_\alpha^\ddag$ of the TS
trajectory. They are
\begin{subequations}
\label{QVars}
\begin{align}
  \left<q_\alpha^{\ddag 2}\right>_\alpha &= \Sigma^2 \;, \\
  \left<v_\alpha^{\ddag 2}\right>_\alpha &= \omega_\text{b}^2\,\Sigma^2 \;, \\
  \left<q_\alpha^\ddag v_\alpha^\ddag\right>_\alpha &= 0
\end{align}
\end{subequations}
with
\begin{equation}
  \label{SigmaDef}
  \Sigma=\frac{g^2}{2\omega_\text{b}^2\sqrt{\gamma^2+4\omega_\text{b}^2}} \;.
\end{equation}
Somewhat surprisingly, the position and velocity of the TS
trajectory turn out to be statistically independent. The variance of the
position coordinate is shown in Fig.~\ref{fig:QVars} as a function of
$\gamma$. Similar to the behavior of the unstable component, it grows with
increasing noise strength from zero to a maximum value of $\kb
T/\omega_\text{b}^2$.

These geometrical observations lead to a novel interpretation of the
friction dependence of reaction rates: In a well-defined reactive system
there is a potential well with a stationary population of reactants on one
side of the barrier. Only the fraction of the population that lies beyond
the stable manifold of the moving reference point, which is very small for
small friction, will be reactive. As the friction increases, the reference
point moves further down into the reactant well so that a larger portion of
the population can at times become reactive. That effect, however, does not
grow indefinitely as the friction gets large.
Instead, it saturates, so that an
opposing effect takes over: As was noted in section~\ref{ssec:WNnoiseless},
the invariant manifolds of the relative dynamics rotate toward the
coordinate axes for large friction, so that the reactive portion of the
relative phase space shrinks to zero. This effect eventually leads to
turnover and to a decreasing overall reaction rate at very large friction.

\subsection{Stochastic separatrices}
\label{ssec:prob}

The previous construction provides a separation between the dynamics of the
barrier crossing, as represented by the relative dynamics, and its
statistical properties, which are embodied in the stochastic TS
Trajectory. The combination of these two aspects allows one to obtain a
detailed picture of the barrier crossing. As an example, we present the
calculation of the stochastic separatrix near the barrier top, i.e. the
collection of all points in phase space for which the reaction probability
is 50\%.\cite{Martens02}

To this end, pick a point $\vec z$ in phase space as the initial point, at
$t=0$, of a stochastic trajectory. 
This point can be characterized equivalently either
by its coordinates $\vec q(0)$, $\vec v(0)$ or by the
diagonal coordinates $z_\text{s}(0), z_\text{u}(0)$ in the stable and
unstable directions. 
The latter representation together consist of
the unstable degree of freedom ---illustrated in figure~\ref{fig:phase}(a),---
and an arbitrary number of stable transverse coordinates. 
The fate of a trajectory in the remote future is
determined by the relative coordinate $\Delta
z_\text{u}(0)=z_\text{u}(0)-z_{\alpha\text{u}}^\ddag(0)$: A trajectory will
evolve into products as $t\to\infty$ if $\Delta z_{\text{u}}(0)>0$, i.e. if
$z_{\alpha\text{u}}^\ddag(0)<z_\text{u}(0)$. Because
$z_{\alpha\text{u}}^\ddag(0)$ is Gaussian distributed with zero mean and a
variance $\sigma_\text{u}^2$ given by~(\ref{WNUnstCorr}) or, in one degree
of freedom, by~(\ref{UVar}), this probability is easily computed to be
\begin{equation}
  \label{P+}
  P_+=\frac{1}{2}\,\operatorname{erfc}
                   \left(-\frac{z_\text{u}(0)}{\sqrt{2}\,\sigma_\text{u}}
		   \right)
\end{equation}
in terms of the complementary error function \cite{Abramowitz}
\begin{equation}
  \operatorname{erfc}(x)=\frac{2}{\sqrt{\pi}}
         \int_x^\infty \exp(-t^2)\,dt \;.
\end{equation}
Similarly, a trajectory will come from the product side in the remote past
$t\to-\infty$ if $z_{\alpha\text{s}}^\ddag(0)<z_\text{s}(0)$, the
probability of which is
\begin{equation}
  \label{P-}
  P_-=\frac{1}{2}\,\operatorname{erfc}
                   \left(-\frac{z_\text{s}(0)}{\sqrt{2}\,\sigma_\text{s}}
		   \right) \;,
\end{equation}
with the variance $\sigma_\text{s}^2$ given by Eq.~\ref{WNStabCorr}
or Eq.~\ref{SVar}.

Equations \ref{P+} and~\ref{P-} provide a means to identify the
stochastic separatrices, where the probabilities for a trajectory to evolve
into reactants or products are both equal to 50\%.  The stochastic
separatrix between trajectories that evolve into reactants or products in
the future is the subspace $z_\text{u}(0)=0$ 
according to Eq.~\ref{P+}, 
whereas the
separatrix between reactants and products in the past is $z_\text{s}(0)=0$.
The present considerations thus reproduce, in a more quantitative way, the
result stated in Ref.~\onlinecite{Martens02} and at the same time 
generalize it to systems with arbitrarily many degrees of freedom. 
Note, however, that the
stochastic separatrices separate reactive from non-reactive trajectories
only in a probabilistic sense, whereas the moving separatrices introduces
in Sec.~\ref{ssec:WNnoiseless} allow one to determine the fate of a
trajectory with certainty.

\subsection{Sampling the TS trajectory}
\label{ssec:sampling}

The time-correlation functions in Eq.~\ref{WNCorr} completely specify the
probability distribution of the TS trajectory, i.e. the joint distribution
of all components of the trajectory at all different times.  To use the TS
trajectory in numerical work of any kind, it is essential 
to obtain random samples $\vec z_\alpha^\ddag(t)$ of the trajectory from
this distribution. The obvious way to do so, namely to generate an instance
of the fluctuating force $\vec\xi_\alpha(t)$ and then compute the
integrals in Eqs.~\ref{xiStab} and \ref{xiUnst}, would quickly turn out to be
prohibitively laborious, so that a more efficient algorithm must be
devised.

The problem at hand can be states more precisely as follows: 
To sample a representative of the TS trajectory $\vec z_\alpha^\ddag(t)$ with
$L$ equidistant time steps over a time interval $T=L\,\Delta t$, one has to
compute $L+1$ random vectors $\vec z_\alpha^\ddag(0),\vec
z_\alpha^\ddag(\Delta t), \dots, \vec z_\alpha^\ddag(T=L\,\Delta t)$ from a
Gaussian distribution with zero mean and covariances given
by Eq.~\ref{WNCorr}. Phrased this way, the problem is to sample a correlated
Gaussian random variable with $2N(L+1)$ components. This is easy to do if
only a single sample point $\vec z_\alpha^\ddag(0)$ on the TS trajectory is
required, but if the number of sample points is large, it is still a
formidable task.

In the case of white noise, this task is greatly simplified by the
recurrence relation
\begin{equation}
  \label{xiRecurr}
  z_{\alpha j}^\ddag(t+s) = e^{\epsilon_j s} 
    \left( z_{\alpha j}^\ddag(t) + \Delta_{\alpha j}(t,s) \right) \;,
\end{equation}
where
\begin{equation}
  \label{DeltaDef}
  \Delta_{\alpha j}(t,s) = \int_t^{t+s} e^{\epsilon_j(t-\tau)} \xi_j(\tau)
              \, d\tau \;,
\end{equation}
that can easily be derived from either Eq.~\ref{xiStab} or Eq.~\ref{xiUnst} 
and holds for both the stable and the unstable components of $\vec
z_\alpha^\ddag(t)$. As the fluctuating force $\xi_{\alpha j}(t)$, the
increment functions $\Delta_{\alpha j}(t,s)$ are Gaussian random variables
with zero mean. Their covariances can be computed 
from Eqs.~\ref{RjWN} and~\ref{WNCorrel} to be
\begin{equation}
  \label{DeltaCov}
  \left<\Delta_{\alpha j}(t,s)\,\Delta_{\alpha k}(t,s)\right>_\alpha =
    \frac{K_{jk}}{\epsilon_j+\epsilon_k}
     \left(1-e^{-(\epsilon_j+\epsilon_k)s}\right) \;,
\end{equation}
where $K_{jk}$ is given by Eq.~\ref{KappaDef}.

Because for white noise the values of the fluctuating
force at different times are independent, the increment functions
$\Delta_{\alpha j}(t_1,s_1)$ and $\Delta_{\alpha k}(t_2,s_2)$ are
stochastically independent if the time intervals $(t_1,t_1+s_1)$ and $(t_2,
t_2+s_2)$ do not overlap, i.e. if $t_1+s_1\le t_2$ or $t_2+s_2\le
t_1$. Further, $\Delta_{\alpha j}(t,s)$ is independent of the component
$z_{\alpha k}^\ddag(\tau)$ of the TS trajectory whenever either $k$ denotes
a stable direction with $\Re\epsilon_k<0$ and $\tau\le t$ or $k$ denotes an
unstable direction with $\Re\epsilon_k>0$ and $\tau\ge t+s$.

These observations lead to the following algorithm for sampling the TS
trajectory: Generate random values for the stable components $z_{\alpha
j}^\ddag(0)$ at time zero using the correlation
function, Eq.~\ref{WNStabCorr}. Generate values for the unstable component
$z_{\alpha j}^\ddag(T)$ at time $T$. They are independent of the stable
components computed before. Typically, there will be only one stable
component. Generate increment vectors $\vec\Delta_{\alpha}(l\,\Delta
t,\Delta t)$ for $l=0,1,\dots,L-1$ using the
covariances, Eq.~\ref{DeltaCov}. Increments for different $l$ are independent.
Using these increments and the recurrence relation in Eq.~\ref{xiRecurr},
compute the stable components $z_{\alpha j}(l\,\Delta t)$ for
$l=1,2,\dots,L$. This recursion is numerically stable because the
exponential factor in Eq.~\ref{xiRecurr} has an absolute value less than
one. The unstable components $z_{\alpha j}(l\,\Delta t)$ are computed using
the same recurrence relation, Eq.~\ref{xiRecurr}, in the opposite order,
starting from $l=L-1,\dots,0$. Again, the recursion is numerically stable.
This algorithm allows one to sample an arbitrary stretch of the TS
trajectory without ever having to compute correlated random variables of a
dimension larger than the phase space dimension $2N$.

\section{The transition state in colored noise}
\label{sec:Coloured}

It might seem at first sight that in the derivation of the TS trajectory in
section~\ref{ssec:WNref} we have never used the assumption that the
fluctuating force $\vec\xi_\alpha(t)$ represents white noise and that,
therefore, the results in Eqs.~\ref{xiStab} and~\ref{xiUnst} for $\vec
z^\ddag_\alpha(t)$ are equally valid if the noise is colored. However, if
the values of the fluctuating force $\xi_\alpha(t)$ at different times are
statistically correlated, the fluctuation-dissipation theorem demands the
occurrence of memory friction effects, and the Langevin
equation, Eq.~\ref{linLE}, must be replaced with the generalized Langevin
equation \cite{zwan61a,zwan61b,prig61,mori65,fox78,grot80,rmp90,zwan01}
\begin{equation}
  \label{linGLE}
  \ddot{\vec q}_\alpha(t) = {\bm\Omega}\vec q_\alpha(t) 
      - \int_{-\infty}^t d\tau\,\bm{\Gamma}(t-\tau)\dot{\vec q}_\alpha(\tau)
      + \vec\xi_\alpha(t) \;,
\end{equation}
where the symmetric-matrix valued function $\bm{\Gamma}(t)$ is related to the
fluctuating force by
\begin{equation}
  \label{FDTColour}
  \left<\vec\xi_\alpha(t)\vec\xi_\alpha^\text{T}(s)\right>_\alpha =
  \kb T\, \bm{\Gamma}(|t-s|) \;.
\end{equation}
Defining
\begin{equation}
  \label{gammaNeg}
  \bm{\Gamma}(t)=0 \text{ if } t<0,
\end{equation}
the upper limit of integration in Eq.~\ref{linGLE} can be 
extended to infinity.

Because the friction force in Eq.~\ref{linGLE} depends on the entire
prehistory of the trajectory rather than only the present velocity,
Eq.~\ref{linGLE} cannot be solved as an initial value
problem. Instead, the entire trajectory up to the present time must be
specified before its further evolution can be determined. Although this
freedom seems to render the phase space of Eq.~\ref{linGLE}
infinite-dimensional, we will show below that in many cases the phase space
dimension is actually finite because only those initial trajectories can be
prescribed that themselves satisfy Eq.~\ref{linGLE}.

To solve the GLE, Eq.~\ref{linGLE}, we use the
bilateral Laplace transform,
\begin{equation}
  \label{Laplace}
  \hat{\vec q}(\epsilon) =
      \int_{-\infty}^\infty dt\, e^{-\epsilon t} \vec q(t) \;,
\end{equation}
and its inverse,\cite{Widder46}
\begin{equation}
  \label{InvLaplace}
  \vec q(t) = \frac{1}{2\pi i}
     \int_{-i\infty}^{i\infty} d\epsilon\, e^{\epsilon t}
                \hat{\vec q}(\epsilon) \;.
\end{equation}
It transforms Eq.~\ref{linGLE} into
\begin{equation}
  \label{ltGLE}
  \left(\epsilon^2+\epsilon \hat{\bm{\Gamma}}(\epsilon)-\bm{\Omega}\right)
       \hat{\vec q}_\alpha(\epsilon) = \hat{\vec\xi}_\alpha(\epsilon) \;.
\end{equation}
Using Eq.~\ref{gammaNeg}, the bilateral Laplace transform
$\hat{\bm{\Gamma}}$ of the friction kernel coincides with the customary
one-sided Laplace transform and is holomorphic on a right half plane of the
complex $\epsilon$-plane that, for bounded $\bm{\Gamma}(t)$, includes the
half plane $\Re\epsilon>0$. We further assume $\hat{\bm{\Gamma}}(\epsilon)$
to be meromorphic on the entire complex plane.

\subsection{Relative dynamics}
\label{ssec:noiselessColour}

The relative coordinate $\Delta\vec q(t)$, which is again given
by Eq.~\ref{relCoord}, satisfies the noiseless version of
Eq.~\ref{ltGLE}, viz.
\begin{equation}
  \label{homGLE}
  \left(\epsilon^2+\epsilon \hat{\bm{\Gamma}}(\epsilon)-\bm{\Omega}\right)
       \Delta\hat{\vec q}(\epsilon) = 0 \;.
\end{equation}
The general solution of Eq.~\ref{homGLE} reads
\begin{equation}
  \label{ltSol}
  \Delta\hat{\vec q}(\epsilon)=\sum_j c_j \vec v_j\,\delta(\epsilon-\epsilon_j)
  \;,
\end{equation}
corresponding to
\begin{equation}
  \label{homSol}
  \Delta\vec q(t) = \sum_j c_j \vec v_j e^{\epsilon_j t} \;,
\end{equation}
where $c_j$ are arbitrary constants and $\epsilon_j, \vec v_j$ denote a
complete linear independent set of solutions of the nonlinear eigenvalue
equation
\begin{equation}
  \label{evalEq}
  \left(\epsilon_j^2+\epsilon_j \hat{\bm{\Gamma}}(\epsilon_j)
       -\bm{\Omega}\right)
      \vec v_j=0 \;, \qquad \vec v_j^2=1 \;.
\end{equation}
Because the friction kernel $\bm{\Gamma}(t)$ is real, its Laplace transform
satisfies
$\hat{\bm{\Gamma}}(\epsilon^\ast)=\hat{\bm{\Gamma}}(\epsilon)^\ast$, where
the asterisk denotes complex conjugation, so that the eigenvalues and
eigenvectors of Eq.~\ref{evalEq} are either real or occur in complex
conjugate pairs. In the latter case, the constants $c_j$ must also be
chosen in conjugate pairs to make $\Delta\vec q(t)$ real. The
solution, Eq.~\ref{homSol}, of the relative dynamics is exactly analogous to
the solution, Eq.~\ref{homSolWhite}, in the case of white noise, except that
the eigenvalues and eigenvectors are computed differently. The construction
of invariant manifolds and a no-recrossing surface that we have given in
section~\ref{ssec:WNnoiseless} therefore carries over unchanged to colored
noise.  A trajectory is specified by prescribing the constants $c_j$.
In other words, the dimension of the phase space of the generalized Langevin
equation, Eq.~\ref{linGLE}, is the number of solutions of the nonlinear
eigenvalue Eq.~\ref{evalEq}. That number is typically larger than
the phase space dimension $2N$ of the memoryless Langevin
equation, Eq.~\ref{linLE}, but still finite if the friction kernel decays 
exponentially. 
The phase space is described implicitly by Eq.~\ref{homSol}.

\subsection{The Transition State Trajectory}
\label{ssec:refColour}

A particular solution of Eq.~\ref{ltGLE} in the presence of the fluctuating
force can be found by solving for $\hat{\vec q}(\epsilon)$
using the inverse Laplace transform~(\ref{InvLaplace}):
\begin{equation}
  \label{refColour}
  \vec q_\alpha^\ddag(t)=\frac{1}{2\pi i} \int_{-i\infty}^{i\infty} d\epsilon\,
     e^{\epsilon t}
     \left(\epsilon^2+\epsilon \hat{\bm{\Gamma}}(\epsilon)
          -\bm{\Omega}\right)^{-1}
     \hat{\vec\xi}_\alpha(\epsilon) \;.
\end{equation}
With the definition of $\hat{\vec\xi}(\epsilon)$ and after the
order of integration is interchanged, Eq.~\ref{refColour} takes the form
\begin{align}
  \label{xiInt}
  \vec q_\alpha^\ddag(t) = \int_{-\infty}^{\infty}d\tau
    &\left[ \frac{1}{2\pi i}\int_{-i\infty}^{i\infty} d\epsilon\,
        e^{\epsilon (t-\tau)}\right.
     \nonumber\\
    &\quad\left.
        \left(\epsilon^2+\epsilon \hat{\bm{\Gamma}}(\epsilon)-\bm{\Omega}
        \right)^{-1}
    \right] \vec\xi_\alpha(\tau) \;.
\end{align}
The integral over $\epsilon$ can be evaluated using the calculus of
residues if the contour of integration is closed through the left half plane
if $\tau<t$ and through the right half plane if $\tau>t$. The poles of the
integrand are precisely the eigenvalues $\epsilon_j$, where the matrix
of Eq.~\ref{evalEq} is not invertible. 
The residue of the matrix inverse in~(\ref{xiInt}) 
at $\epsilon=\epsilon_j$ is denoted by $\bm{\mu}_j$, so that
\begin{equation}
  \label{xiColour}
  \vec q_\alpha^\ddag(t)=\sum_j \vec q^\ddag_{\alpha j}(t)
\end{equation}
with
\begin{align}
  \label{xiStabColour}
    \vec q^\ddag_{\alpha j}(t) &= \bm{\mu}_j \int_{-\infty}^t d\tau\,
        e^{\epsilon_j(t-\tau)} \vec\xi_\alpha(\tau) \nonumber\\
      &= \bm{\mu}_j \int_{-\infty}^0 d\tau\,e^{-\epsilon_j\tau}
	       \vec\xi_\alpha(t+\tau)
\end{align}
if $\Re\epsilon_j<0$ and
\begin{align}
  \label{xiUnstColour}
    \vec q^\ddag_{\alpha j}(t) &= -\bm{\mu}_j \int_t^\infty d\tau\,
         e^{\epsilon_j(t-\tau)} \vec\xi_\alpha(\tau) \nonumber\\
      &= -\bm{\mu}_j\int_0^\infty d\tau\,e^{-\epsilon_j\tau} \vec\xi_\alpha(t+\tau)
\end{align}
if $\Re\epsilon_j>0$. These solutions are close analogues of
Eqs.~\ref{xiStab} and \ref{xiUnst} 
in the case of white noise, with the residue factors
$\bm{\mu}_j$ replacing the projection onto the eigenspace $j$. The components
$\vec q^\ddag_{\alpha j}$ can easily be verified to satisfy the equation of
motion
\begin{equation}
  \label{xiEq}
  \dot{\vec q}^\ddag_{\alpha j}(t) 
     = \epsilon_j \vec q^\ddag_{\alpha j}(t) + \bm{\mu}_j\vec\xi_\alpha(t)
\end{equation}
that is expected for a component in an eigenspace of the noiseless dynamics
with eigenvalue $\epsilon_j$. Again, the residues $\bm{\mu}_j$ play the role of
the projection onto the eigenspace $j$.

A more explicit form of the residues $\bm{\mu}_j$ can be found 
as follows:
First, by observing that
\begin{equation}
  \label{AEpsDef}
  \bm{A}(\epsilon) = \epsilon^2+\epsilon \hat{\bm{\Gamma}}(\epsilon)
                    -\bm{\Omega}
\end{equation}
is a symmetric matrix, we immediately conclude that
$\bm{\mu}_j$ (=$\bm{\mu}_j^\text{T}$)
is also symmetric. 
For any fixed $\epsilon$, the matrix $\bm{A}(\epsilon)$
possesses a complete
set of eigenvalues $\eta_i(\epsilon)$ and orthonormal eigenvectors $\vec
w_i(\epsilon)$:
\begin{align}
  \label{AEpsEval}
  \bm{A}(\epsilon)\vec w_i(\epsilon) = \eta_i(\epsilon) \vec w_i(\epsilon)\;,
  \nonumber\\
  \vec w_i(\epsilon)\cdot \vec w_j(\epsilon) = \delta_{ij} \;.
\end{align}
In terms of its eigensystem, the matrix $\bm{A}(\epsilon)$ can be written as
\begin{equation}
  \bm{A}(\epsilon) = \sum_i \eta_i(\epsilon) \vec w_i(\epsilon)
     \vec w_i^\text{T}(\epsilon) \;,
\end{equation}
so that
\begin{equation}
  \label{AInv}
  \bm{A}^{-1}(\epsilon) = \sum_i \frac{1}{\eta_i(\epsilon)} \vec w_i(\epsilon)
     \vec w_i^\text{T}(\epsilon) \;.
\end{equation}
When $\epsilon$ is equal to a solution $\epsilon_j$ of the nonlinear
eigenvalue problem in Eq.~\ref{evalEq}, one of the eigenvalues $\eta_i$
vanishes, say $\eta_1(\epsilon_j)=0$, and $\vec w_1(\epsilon_j)=\vec
v_j$. If all the eigenvalues are simple, only the term $i=1$
in Eq.~\ref{AInv} has a pole at $\epsilon=\epsilon_j$, and its residue is
\begin{equation}
  \label{muProj}
  \bm{\mu}_j=\kappa_j \vec v_j\vec v_j^\text{T} \;,
\end{equation}
where
\begin{equation}
  \kappa_j=\frac{1}{\eta'(\epsilon_j)}
\end{equation}
and the prime denotes the derivative with respect to $\epsilon$. The
residue $\bm{\mu}_j$ is thus proportional to the geometric projection onto the
eigenvector $\vec v_j$, with the factor $\kappa_j$ describing the strength
of the coupling of the noise to the particular eigenmode. Using first-order
perturbation theory, this factor is found to be
\begin{equation}
  \kappa_j=\frac{1}{2\epsilon_j+\vec v_j^\text{T}
  \left(\hat{\bm{\Gamma}}(\epsilon_j)+
        \epsilon_j\hat{\bm{\Gamma}}'(\epsilon_j)\right)
  \vec v_j} \;.
\end{equation}

The representation, Eq.~\ref{muProj}, of the residues leads to a more explicit
form of the components $q^\ddag_{\alpha j}(t)$:
\begin{equation}
  \vec q^\ddag_{\alpha j}(t) = \kappa_j\vec v_j
      \int_{-\infty}^0 d\tau\,e^{-\epsilon_j\tau}
                \vec v_j\cdot \vec\xi_\alpha(t+\tau)
\end{equation}
if $\Re\epsilon_j<0$, and a similar expression for $\Re\epsilon_j>0$. Note
that $\vec q^\ddag_{\alpha j}(t)$ is a scalar multiple of the
eigenvector $\vec v_j$.

The eigenvectors of the linear eigenvalue problem in Eq.~\ref{AEpsEval}
satisfy the completeness condition
\begin{equation}
  \label{linComplete}
  \sum_i \vec w_i\vec w_i^\text{T} = {\bm I} \;,
\end{equation}
where $\bm I$ is the $N\times N$ unit matrix. Because the eigenvalue
problem in Eq.~\ref{evalEq} is nonlinear, the analogous relation for the $\vec
v_j$ does not hold. Instead, if $\hat{\bm{\Gamma}}(\epsilon)$ is
meromorphic at infinity, the residues $\bm{\mu}_j$ satisfy the completeness
conditions
\begin{subequations}
\label{muComplete}
\begin{gather}
  \label{muComplete1}
  \sum_j \bm{\mu}_j = \sum_j \kappa_j\vec v_j \vec v_j^\text{T} = \bm{0} \;, \\
  \label{muComplete2}
  \sum_j \epsilon_j \bm{\mu}_j 
     = \sum_j \kappa_j\epsilon_j\vec v_j \vec v_j^\text{T} = \bm{I} \;.
\end{gather}
\end{subequations}
To see this, note that
\begin{equation}
  \label{muProof1}
  \sum_j \bm{\mu}_j = \frac{1}{2\pi i} \oint_{\cal C} d\epsilon
     \left(\epsilon^2+\epsilon\,\hat{\bm{\Gamma}}(\epsilon)
           -\bm{\Omega}\right)^{-1}
  \;,
\end{equation}
where $\cal C$ is a contour in the complex $\epsilon$-plane that encircles
all eigenvalues $\epsilon_j$ in the clockwise direction. With
$\theta=1/\epsilon$, Eq.~\ref{muProof1} reads
\begin{equation}
  \sum_j \bm{\mu}_j = \frac{1}{2\pi i} \oint_{1/\cal C} d\theta 
     \left({\bm I}+\theta\,\hat{\bm{\Gamma}}(1/\theta)
           -\theta^2\bm{\Omega}\right)^{-1}
   \;,
\end{equation}
where the contour $1/\cal C$ is again oriented clockwise. By construction,
it encircles none of the inverse eigenvalues $1/\epsilon_j$. As the
integrand is furthermore regular at $\theta=0$, the integral is zero, thus
proving Eq.~\ref{muComplete1}. To prove Eq.~\ref{muComplete2}, a 
similar computation leads to
\begin{align}
\label{muProof2}
  \sum_j \epsilon_j\bm{\mu}_j &= 
    \frac{1}{2\pi i} \oint_{\cal C} d\epsilon\, \epsilon
     \left(\epsilon^2+\epsilon \,\hat{\bm{\Gamma}}(\epsilon)
           -\bm{\Omega}\right)^{-1} \nonumber\\
   &= \frac{1}{2\pi i} \oint_{1/\cal C} \frac{d\theta}{\theta} \,
      \left({\bm I}+\theta\,\hat{\bm{\Gamma}}(1/\theta)
            -\theta^2{\bm\Omega}\right)^{-1} \nonumber\\
   &= {\bm I} \;,
\end{align}
where the last equality holds because
the integrand in Eq.~\ref{muProof2} has a simple pole at $\theta=0$.

\subsection{Statistical properties of the TS trajectory}
\label{ssec:statColour}

As in the case of white noise, the components of the TS
trajectory,
Eqs.~\ref{xiStabColour} and~\ref{xiUnstColour}, are Gaussian
distributed with zero mean if the noise $\vec\xi_\alpha(t)$ is. To specify
their distribution completely, it remains only to compute their time
correlation functions. 
For $t>0$ and in the case of two stable components
with $\Re\epsilon_j<0$, $\Re\epsilon_k<0$, they are
\begin{align}
  \label{ColourCorr}
    \left<\vec q^\ddag_{\alpha j}(t)
          \vec q^{\ddag\text{T}}_{\alpha k}(0)\right>_\alpha &=
      \bm{\mu}_j\,\int_{-\infty}^t d\tau \int_{-\infty}^0 d\sigma\,
         e^{-\epsilon_j(\tau-t)} e^{-\epsilon_k\sigma}
    \nonumber\\&\quad\quad\quad
       \left<\vec\xi_\alpha(\tau)\vec\xi_\alpha(\sigma)\right>_\alpha
       \bm{\mu}_k^\text{T}
       \nonumber\\
     &= \kb T \,\bm{\mu}_j\, \bm{I}_{jk}(t)\, \bm{\mu}_k
\end{align}
with the matrix-valued correlation integral
\begin{equation}
  \label{ISSDef}
  \bm{I}_{jk}(t) = \int_{-\infty}^t d\tau \int_{-\infty}^0 d\sigma\,
      e^{\epsilon_j(t-\tau)} e^{-\epsilon_k\sigma}\bm{\Gamma}(|\tau-\sigma|)
  \;.
\end{equation}
Recall that $\bm{\mu}_k$ is symmetric, 
and note that the time correlation matrix in Eq.~\ref{ColourCorr} 
is a scalar multiple of
the matrix $\vec v_j\vec v_k^\text{T}$
if Eq.~\ref{muProj} holds.

To evaluate the correlation integral in 
Eq.~\ref{ISSDef}, we split the range of
the $\tau$ integration into the intervals $(-\infty,0)$ and $(0,t)$. The
latter part can then be expressed through the function
\begin{equation}
  \label{CDef}
  \bm{C}(\zeta_1,\zeta_2,t) = \int_0^t d\tau \int_0^{\infty} d\sigma\,
     e^{\zeta_1(t-\tau)} e^{-\zeta_2\sigma} \bm{\Gamma}(\tau+\sigma) \;.
\end{equation}
The former part, apart from a factor $e^{\epsilon_j t}$, is
\begin{equation}
  \label{ISS1}
  \bm{I}_{jk}(0) =   \int_{-\infty}^0 d\tau \int_{-\infty}^0 d\sigma\,
      e^{-\epsilon_j\tau} e^{-\epsilon_k\sigma} \bm{\Gamma}(|\tau-\sigma|) \;,
\end{equation}
which in terms of $\chi=\tau-\sigma$ and $\psi=\tau+\sigma$, can be
rewritten as
\begin{align}
    \bm{I}_{jk}(0) &= \frac 12
      \int_{-\infty}^\infty \!d\chi\, \bm{\Gamma}(|\chi|)\,
          e^{(\epsilon_k-\epsilon_j)\chi/2}
      \int_{-\infty}^{|\chi|} d\psi\,
          e^{-(\epsilon_j+\epsilon_k)\psi/2} \nonumber\\
    &= \frac{1}{-(\epsilon_j+\epsilon_k)} \left(
           \int_{-\infty}^0 d\chi\, e^{\epsilon_k\chi}\, \bm{\Gamma}(-\chi) 
        \right.\nonumber\\
    &\text{\phantom{$\frac{1}{-(\epsilon_j+\epsilon_k)}$}}\quad
           \left.
             + \int_0^\infty d\chi\, e^{-\epsilon_j\chi}\, \bm{\Gamma}(\chi)
           \right) \nonumber\\
    &= \frac{1}{-(\epsilon_j+\epsilon_k)}
           \left(
             \hat{\bm{\Gamma}}(-\epsilon_j) + \hat{\bm{\Gamma}}(-\epsilon_k)
           \right) \;.
\end{align}
The correlation integral in Eq.~\ref{ISSDef} thus finally reads
\begin{subequations}
\begin{align}
  \bm{I}_{jk}(t) &= \frac{e^{\epsilon_jt}}{-(\epsilon_j+\epsilon_k)}
           \left(
             \hat{\bm{\Gamma}}(-\epsilon_j) + \hat{\bm{\Gamma}}(-\epsilon_k)
           \right)
           + \bm{C}(\epsilon_j,-\epsilon_k,t) \nonumber\\
    &\qquad\qquad\text{if $\Re\epsilon_j<0$, $\Re\epsilon_k<0$}\;.\\
\intertext{The correlation functions involving unstable components take
           the same general form of Eq.~\ref{ColourCorr} with}
  \bm{I}_{jk}(t) &= \frac{e^{-\epsilon_kt}}{\epsilon_j+\epsilon_k}
           \left(
             \hat{\bm{\Gamma}}(\epsilon_j) + \hat{\bm{\Gamma}}(\epsilon_k)
           \right)
           + \bm{C}(-\epsilon_k,\epsilon_j,t) \nonumber\\
    & \qquad\qquad\text{if $\Re\epsilon_j>0$, $\Re\epsilon_k>0$}\;,
      \label{IUU}\\ 
  \bm{I}_{jk}(t) &= \frac{e^{-\epsilon_kt}}{\epsilon_j+\epsilon_k}
           \left(
             \hat{\bm{\Gamma}}(\epsilon_k) + \hat{\bm{\Gamma}}(-\epsilon_k)
           \right)\nonumber\\
         &-\frac{e^{\epsilon_jt}}{\epsilon_j+\epsilon_k}
           \left(
             \hat{\bm{\Gamma}}(-\epsilon_j) + \hat{\bm{\Gamma}}(-\epsilon_k)
           \right)
           + \bm{C}(\epsilon_j,-\epsilon_k,t) \nonumber\\
    & \qquad\qquad\text{if $\Re\epsilon_j<0$, $\Re\epsilon_k>0$}\;,\\ 
  \bm{I}_{jk}(t) &= \frac{e^{\epsilon_jt}}{\epsilon_j+\epsilon_k}
           \left(
             \hat{\bm{\Gamma}}(\epsilon_j) - \hat{\bm{\Gamma}}(-\epsilon_k)
           \right)
           + \bm{C}(\epsilon_j,-\epsilon_k,t) \nonumber\\
    & \qquad\qquad\text{if $\Re\epsilon_j>0$, $\Re\epsilon_k<0$}\;.
\end{align}
\end{subequations}

All correlation integrals consist of exponentially decaying contributions
and a correction term $\bm{C}$. The latter can be written as
\begin{equation}
  \bm{C}(\zeta_1,\zeta_2,t) = \int_0^t d\tau\,
    e^{\zeta_1(t-\tau)}\, \hat{\bm{\Gamma}}_\tau(\zeta_2) \;,
\end{equation}
where $\hat{\bm{\Gamma}}_\tau$ denotes the Laplace transform of the
time-shifted function $\bm{\Gamma}(t+\tau)$. Unless
$\hat{\bm{\Gamma}}(\epsilon)$ is growing at least exponentially as
$|\epsilon|\to\infty$ and $\Re\epsilon<0$, an explicit expression for the
correction term $\bm{C}$ can be found. To this end, use the definition
of $\hat{\bm{\Gamma}}_\tau$ and the inverse Laplace
transform, Eq.~\ref{InvLaplace}, to write, for $\Re\epsilon>0$,
\begin{align}
  \label{GammaTau}
  \hat{\bm{\Gamma}}_\tau(\epsilon) &= \int_0^\infty d\sigma\,
     e^{-\epsilon\sigma}\,
     \frac{1}{2\pi i}\int_{-i\infty}^{i\infty}d\eta\, e^{\eta(\sigma+\tau)}\,
     \hat{\bm{\Gamma}}(\eta) \nonumber\\
   &= \frac{1}{2\pi i}\int_{-i\infty}^{i\infty}d\eta\, e^{\eta\tau}
      \frac{\hat{\bm{\Gamma}}(\eta)}{\epsilon-\eta} \;.
\end{align}
If it is now permissible to close the integration contour through the left
half plane, Eq.~\ref{GammaTau} yields
\begin{equation}
  \hat{\bm{\Gamma}}_\tau(\epsilon) = \sum_i \frac{\nu_i}{\epsilon-E_i}\,
      e^{E_i\tau} \;,
\end{equation}
where $E_i$ are the poles of $\hat{\bm{\Gamma}}(\epsilon)$ in the left half
plane (which typically are all poles) and $\nu_i$ are the corresponding
residues. The correction term $\bm{C}$ thus reads
\begin{equation}
  \label{CResidue}
  \bm{C}(\zeta_1,\zeta_2,t)=\sum_i
    \frac{\nu_i}{(E_i-\zeta_1)(E_i-\zeta_2)}
    \left( e^{\zeta_1t}-e^{E_it}\right)
\end{equation}
and is also exponentially decaying.

\subsection{Reconstructing the eigenvalues}
\label{sssec:xiLT}

The dynamics of the GLE, Eq.~\ref{linGLE}, is
characterized by the eigenvalues $\epsilon_j$. 
If the TS trajectory does indeed capture the essential
aspects of the dynamics, 
it should enable us to identify the eigenvalues.
In particular, we should be able to find
the Grote-Hynes reaction frequency \cite{grot80,pollak86b} viz.\ the unique
eigenvalue with positive real part, that approximately describes the
influence of colored noise on the reaction rate.  
That this information can actually be obtained from 
the TS trajectory can be seen most easily by examining
the statistical properties of its Laplace transform 
rather than the time-correlation functions of the TS trajectory directly.
According to
Eq.~\ref{refColour},
\begin{equation}
  \label{xiLT}
  \hat{\vec q}_\alpha^\ddag(\epsilon) = 
      \left(\epsilon^2+\epsilon \hat{\bm{\Gamma}}(\epsilon)
           -\bm{\Omega}\right)^{-1}
       \hat{\vec\xi}_\alpha(\epsilon) \;.
\end{equation}
This Laplace transform has poles at the eigenvalues $\epsilon_j$. In
addition, there are the random poles of the Laplace transform
$\hat{\vec\xi}_\alpha(\epsilon)$ of the noise. Thus, the eigenvalues of the
noiseless dynamics cannot be determined from the Laplace transform of a
single instance of the TS trajectory. If, however, the Laplace transform of
several instances of $\vec q^\ddag_\alpha$ is given, the eigenvalues can be
discerned with certainty because they are fixed whereas the poles of
$\hat{\vec\xi}_\alpha(\epsilon)$ are random.

The randomness of the poles is entirely absent from the correlation
function of $\hat{\vec q}_\alpha^\ddag$, which is
\begin{align}
  \label{xiLtCorr}
  \left<\hat{\vec q}_\alpha^\ddag(\epsilon)\,
        \hat{\vec q}_\alpha^{\ddag\text{T}}(\epsilon')\right>_\alpha
  =& \left(\epsilon^2+\epsilon \hat{\bm{\Gamma}}(\epsilon)-\Omega\right)^{-1}
    \left<\hat{\vec\xi}_\alpha(\epsilon)
          \hat{\vec\xi}_\alpha^\text{T}(\epsilon')\right>_\alpha
    \nonumber\\ 
  &\times
    \left(\epsilon'^2+\epsilon' \hat{\bm{\Gamma}}(\epsilon')-\Omega\right)^{-1}
  \;.
\end{align}
If the Laplace transform $\hat{\vec\xi}_\alpha(\epsilon)$ is replaced with
its definition in Eq.~\ref{Laplace}, the remaining average turns out to be
\begin{equation}
  \label{xiLapAv}
  \left<\hat{\vec\xi}_\alpha(\epsilon)
         \hat{\vec\xi}_\alpha^\text{T}(\epsilon')\right>_\alpha
  = \int_0^\infty d\tau \int_0^\infty d\sigma \,
        e^{-\epsilon\tau} e^{-\epsilon'\sigma}\,
        \bm{\Gamma}(|\tau-\sigma|) \;.
\end{equation}
In the notation of Sec.~\ref{ssec:statColour}, Eq.~\ref{xiLapAv}
is the correlation integral $\bm{I}_{\epsilon\epsilon'}(0)$ between two
unstable components. It has
already been evaluated in Eq.~\ref{IUU} to be
\begin{equation}
   \left<\hat{\vec\xi}_\alpha(\epsilon)
         \hat{\vec\xi}_\alpha^\text{T}(\epsilon')\right>_\alpha
  = \frac{\hat{\bm{\Gamma}}(\epsilon)+\hat{\bm{\Gamma}}(\epsilon')}
         {\epsilon+\epsilon'}\;.
\end{equation}
Therefore,
\begin{align}
   \left<\hat{\vec q}_\alpha^\ddag(\epsilon)\,
         \hat{\vec q}_\alpha^{\ddag^\text{T}}(\epsilon')\right>_\alpha
  =& \left(\epsilon^2+\epsilon \hat{\bm{\Gamma}}(\epsilon)-\Omega\right)^{-1}
    \frac{\hat{\bm{\Gamma}}(\epsilon)+\hat{\bm{\Gamma}}(\epsilon')}
         {\epsilon+\epsilon'}\nonumber\\
   &\times
    \left(\epsilon'^2+\epsilon' \hat{\bm{\Gamma}}(\epsilon')-\Omega\right)^{-1}
\end{align}
and in particular, for $\epsilon=\epsilon'$,
\begin{align}
   \label{xiLTCorr}
     \left<\hat{\vec q}_\alpha^\ddag(\epsilon)\,
           \hat{\vec q}_\alpha^{\ddag\text{T}}(\epsilon)\right>_\alpha
    =& \left(\epsilon^2+\epsilon \hat{\bm{\Gamma}}(\epsilon)-\Omega\right)^{-1}
             \frac{\hat{\bm{\Gamma}}(\epsilon)}{\epsilon} \nonumber\\
    &\times
       \left(\epsilon^2+\epsilon \hat{\bm{\Gamma}}(\epsilon)-\Omega
       \right)^{-1} \;.
\end{align}
This correlation function has poles at $\epsilon=0$, at the poles of
$\hat{\bm{\Gamma}}(\epsilon)$ and at the eigenvalues of the noiseless
dynamics. Usually, the friction kernel $\bm{\Gamma}$ will be known, so that the
eigenvalues $\epsilon_j$ can be determined from
Eq.~\ref{xiLTCorr}. On the
other hand, even if the friction kernel is unknown, the eigenvalues can be
discerned because they are double poles of
Eq.~\ref{xiLTCorr}, whereas the
poles of $\hat{\bm{\Gamma}}(\epsilon)$ will typically be simple. In particular,
the Grote-Hynes frequency \cite{grot80,pollak86b} 
can be identified as the only pole of
the correlation function in Eq.~\ref{xiLTCorr} that has a positive real part.

\section{Conclusions and outlook} 

To generalize the formalism of TST to reactive systems
driven by noise, we have introduced a time-dependent dividing surface that
is stochastically moving in phase space so that it is crossed once and only once
by each transition path.  
This construction requires the identification of a privileged
stochastic trajectory that remains in the vicinity of the barrier for all
times. 
This TS Trajectory serves as the origin of a stochastically
moving coordinate system that carries the no-recrossing TS
dividing surface for the noiseless dynamics. 
The moving no-recrossing dividing surface thus obtained describes 
a reaction in the same way 
as a static dividing surface in conventional TST 
and therefore provides a novel approach to the theory of reaction dynamics
in a noisy environment.
The challenge lying ahead is to determine a rate formula that makes
optimal use of the geometric objects associated with 
the TS trajectory and the moving no-recrossing TS dividing surface.

The further generalization of this approach to all-atom representations 
of solvated chemical reactions is nontrivial.  
In the Langevin models,
the noise sequence can be specified independently of 
the reaction path or the solvent modes.
For an all-atom solvent, by contrast, this is impossible
because the noise sequence depends on the trajectory of the
chosen coordinates identifying the reaction:
A given solvation-shell structure creates an effective cavity
that accomodates a particular set of reaction geometries
more easily than others.
Thus a critical question to be pursued in future work is how to identify
the reaction geometry of the moving TS dividing surface in
all-atom representations of a solvated chemical reaction.

\begin{acknowledgments}
This work was partly supported by the US National
Science Foundation and by the Alexander von Humboldt-Foundation.
\end{acknowledgments}

\end{document}